\documentclass[onecolumn,prb,10pt,aps,longbibliography,superscriptaddress]{revtex4-2}

\usepackage{subcaption}
\usepackage{graphicx}
\usepackage{amsmath}
\usepackage{amssymb}
\usepackage{amsfonts}
\usepackage{dsfont}
\usepackage{dcolumn}
\usepackage{bbm}
\usepackage{ragged2e}
\usepackage{mathtools}
\usepackage{braket}
\usepackage{physics}
\usepackage{siunitx}
\usepackage[unicode]{hyperref}
\usepackage[nameinlink]{cleveref}
\Crefname{section}{Sec.}{Secs.}
\Crefname{equation}{Eq.}{Eqs.}
\Crefname{figure}{Fig.}{Figs.}
\Crefname{tabular}{Tab.}{Tabs.}

\usepackage{bbold} % For \mathbb{1}
\usepackage{nicefrac}
\usepackage{tikz}
\usetikzlibrary{arrows.meta}
\usepackage{lipsum}  
\usepackage{xcolor}

\usepackage[makeroom]{cancel}

\usepackage{xfrac}

\usepackage[normalem]{ulem}
\definecolor{tdo_blue}{HTML}{0000FF}
\definecolor{tdo_darkgreen}{HTML}{839A00}
\definecolor{tdo_red}{HTML}{FF0000}

\newcommand{\bes}{\begin{subequations}}
\newcommand{\ees}{\end{subequations}}
\newcommand{\be}{\begin{equation}}
\newcommand{\ee}{\end{equation}}

\makeatletter
  \newcommand\flausr{\@fleqntrue}
\makeatother

\begin{document}
    %==============================================================================
    % Title
    %==============================================================================
    \title{Entanglement-informed Construction of Variational Quantum Circuits}

    %==============================================================================
    % Authors and Affiliation
    %=====================================================

    \author{Alina Joch}
    \email{alina.joch@tu-dortmund.de}%suggestions
    \affiliation{Condensed Matter Theory, TU Dortmund University, Otto-Hahn-Stra\ss{}e 4, 44227 Dortmund, Germany}
    \affiliation{Institute of Software Technology, German Aerospace Center (DLR), 51147 Cologne, Germany}

    \author{Götz S. Uhrig}
    \email{goetz.uhrig@tu-dortmund.de}
    \affiliation{Condensed Matter Theory, TU Dortmund University, Otto-Hahn-Stra\ss{}e 4, 44227 Dortmund, Germany}

    \author{Benedikt Fauseweh}
    \email{benedikt.fauseweh@tu-dortmund.de}
    \affiliation{Condensed Matter Theory, TU Dortmund University, Otto-Hahn-Stra\ss{}e 4, 44227 Dortmund, Germany}
    \affiliation{Institute of Software Technology, German Aerospace Center (DLR), 51147 Cologne, Germany}

    \date{\textrm{\today}}

    %==============================================================================
    % Abstract
    %==============================================================================
    \begin{abstract}
    The Variational Quantum Eigensolver (VQE) is a promising tool for simulating ground states of quantum many-body systems on noisy quantum computers. Its effectiveness relies heavily on the ansatz, which must be both hardware-efficient for implementation on noisy hardware and problem-specific to avoid local minima and convergence problems. In this article, we explore entanglement-informed ansatz schemes that naturally emerge from specific models, aiming to balance accuracy with minimal use of two-qubit entangling gates, allowing for efficient use of techniques such as quantum circuit cutting. We focus on three models of quasi-1D Hamiltonians: (i) systems with impurities acting as entanglement barriers, (ii) systems with competing long-range and short-range interactions transitioning from a long-range singlet to a quantum critical state, and (iii) random quantum critical systems. For the first model, we observe a plateau in the ansatz accuracy, controlled by the number of entangling gates between subsystems. This behavior is explained by iterative capture of  eigenvalues in the entanglement spectrum. In the second model, combining long-range and short-range entanglement schemes yields the best overall accuracy, leading to global convergence in the entanglement spectrum. For the third model, we use an renormalization group approach to build the short- and long-range entanglement structure of the ansatz. Our comprehensive analysis provides a new perspective on the design of ansätze based on the expected entanglement structure of the approximated state. 
		\end{abstract}
		
    \maketitle

    %==============================================================================
    % Introduction  
    %==============================================================================
    \section{Introduction}
    \label{s:introduction}

    Since Feynman proposed using quantum computers to simulate quantum mechanical systems \cite{feynm82}, various technologies have emerged as platforms for quantum simulation, for an overview see Ref.\  \cite{fause24}. However, the development of large-scale, fault-tolerant quantum computers remains a significant challenge. Currently, only noisy intermediate-scale quantum (NISQ) devices \cite{presk18} are available, which necessitates the reduction of quantum circuits to a minimal number of gates and qubits \cite{PhysRevX.5.021027,doi:10.1126/science.1208001,Barends2015,Fauseweh2021}.

    In order to still take advantage of these devices, special algorithms have been developed to fulfill these restrictions \cite{bhart22}. 
    One promising approach is the use of hybrid classical-quantum algorithms, such as variational quantum algorithms (VQAs) \cite{cerez21,bhart22,Kandala2017,Higgott2019variationalquantum,Barison2021efficientquantum,Fauseweh2023quantumcomputing}. VQAs address computational tasks through a combination of classical and quantum computation. In these algorithms, the quantum computation involves a parametrized quantum circuit, the ansatz, that depends on a set of classical parameters, such as the rotation angles of single-qubit gates. These parameters are optimized iteratively through classical computation by minimizing a cost function designed such that its minima correspond to the solution of the computational problem. Through several quantum-classical optimization loops, the optimal parameters are identified.

    One of the most prominent VQAs is the variational quantum eigensolver (VQE) \cite{cerez21, peruz14}, which is used to provide an approximation to the ground state of many-body quantum systems through a unitary transformation of the computational basis state.
    The circuit ansatz typically consists of several layers of the same set of quantum gates.
    
    In principle, the accuracy of the optimized VQE results increases with the number of layers used \cite{bravo20}. Quantum states prepared by parameterized quantum circuits offer enhanced expressivity in terms of captured entanglement compared to low-rank tensor states that can be computed classically \cite{PhysRevResearch.2.033125,https://doi.org/10.1002/qute.201900070}. VQAs, therefore, can be effectively compared to classical variational methods such as tensor networks \cite{Orus2019} and variational Monte Carlo \cite{Clark_2018}.
    
    On NISQ devices increasing the number of layers also introduces higher levels of noise due to the increased depth of the quantum circuit. Additionally, deeper VQE circuits exhibit the barren plateau phenomenon, making it exponentially harder to optimize the circuit due to a vanishing gradient \cite{McClean2018}, even on fault-tolerant quantum computers.
    Therefore, choosing an appropriate ansatz is a key task to enhance the accuracy, noise robustness, and optimizability of VQE \cite{lyu20, tkach21}. At the same time, it is of fundamental interest to understand how different ansatz circuits converge and how the underlying entanglement of the approximated state affects this convergence \cite{bravo20}. 

In this paper, we investigate three distinct models in which an ansatz emerges naturally due to the structure and interaction strength of the terms in the Hamiltonian. In the first case, we investigate systems with impurities that act as entanglement barriers. Here, we aim to reduce the number of gates that entangle the different subsystems separated by the impurities. Such an approach makes a technique called circuit cutting \cite{bravy16,peng20b,mitar21} attractive: The quantum circuit is cut at several points and the smaller sub-circuits are evaluated separately on a quantum computer. The information of the different parts is exchanged classically.  Since the classical information exchange is costly, it is only reasonable to use this method, if few cuts are sufficient in order to be able to reach accurate results. However, in systems where this is possible, circuit cutting significantly reduces the number of quantum gates as well as qubits in a calculation, which can further streamline VQE approaches by decreasing the hardware demands and potentially mitigating errors introduced through gate operations. 

An additional benefit of this approach is that it addresses the problem of barren plateaus. By reducing the expressivity of our ansatz through a reduction of entangling gates, the distance of the circuit to a 2-design is increased, which has recently been connected to the magnitude of the gradient in circuit optimization \cite{PRXQuantum.3.010313}. 

The second and third case focuses on quantum critical systems and systems with long-range entanglement, as they are especially challenging for VQE \cite{lyu23}. We will investigate whether singlets on long-range interacting dimers as an initial state can decrease the number of layers while still achieving a desired accuracy in both cases.

Besides the accuracy in the energy, we also investigate the two approaches through the lens of entanglement spectra. Understanding entanglement properties of strongly correlated systems with and without disorder is of significant interest in the condensed matter community \cite{affle09, proda10, torla14}. Additionally, it is also started to be used to understand the behavior of quantum circuits in VQA \cite{wiers20, bravo20, sun23}.

    The article is set up as follows. The studied models are 
    motivated and discussed in Sec.\@~\ref{s:model}. 
    Then, Sec.\@~\ref{s:methods} explains the used methods including a more detailed 
    description of the chosen ansatz schemes for the VQE.
    Subsequently, in Sec.\@~\ref{s:impurity} we investigate the accuracy of the ground state energy, entanglement entropy and spectrum for the impurity model.
    Then, in Sec.\@~\ref{s:singletstate} we discuss the energy and entanglement spectrum of the long-range Heisenberg chain for different ansatz schemes.
    In Sec.\@~\ref{s:random_model}, we focus on a model with a 
    random quantum critical point.
    Finally, we conclude in Sec.\@~\ref{s:conclusion}.

	  %==============================================================================
    % Model
    %==============================================================================
    \section{Models}
    \label{s:model}  

    \subsection{Local impurity model}

\begin{figure}[!ht]
      \centering
      \includegraphics[width=5.5cm]{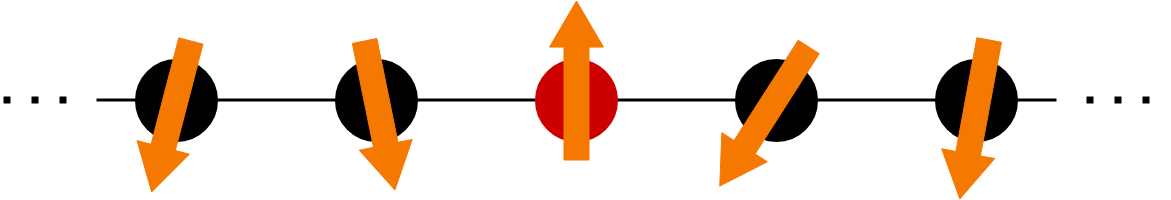}
      \vspace{-0.5em}
      \caption[]
			{\justifying The impurity model is depicted. The qubit in the center marked in red is subject to a strong longitudinal magnetic field.}
      \label{f:impuritymodel}
    \end{figure}

    The first model we investigate is the analytically solvable transverse field Ising model (TFIM) \cite{tfim, tfim2} with a central impurity qubit. 
    Its Hamiltonian is given by 
    \begin{equation}\label{tfim}
      H_{\mathrm{TFIM}} = \sum_{i=-L/2}^{L/2} \left( - \sigma_i^z \sigma_{i+1}^z + h^x \sigma_i^x \right) + h_0^z \sigma_0^z \, , %J=-1
    \end{equation}
    where $h^x$ is the disordering magnetic field.
    The last term describes a longitudinal field at the central qubit of the chain. Without longitudinal field and for $| h^x | < 1$, the system is in a ferromagnetic phase with
    all qubits being aligned along the $z$ direction for $| h^x | \rightarrow 0$. 
    For $| h^x | > 1$, the system is in a paramagnetic phase. In the limit of $| h^x |\rightarrow \infty$, all qubits are aligned along the $x$ direction.
    Between the two phases, at $| h^x | = 1$, there is a quantum critical point, where the system is gapless.
    In our studies, we will focus on the critical point with $h^x = -1$. The model is visualized in 
    Fig.\@ \ref{f:impuritymodel}.

    We calculate the ground state energy numerically by diagonalization making it possible to 
    compare these results with the results obtained with VQE.
    We will in all cases use open boundary conditions.

    \subsection{Long-range entanglement model}

\begin{figure}[!ht]
      \centering
      \includegraphics[width=9.0cm]{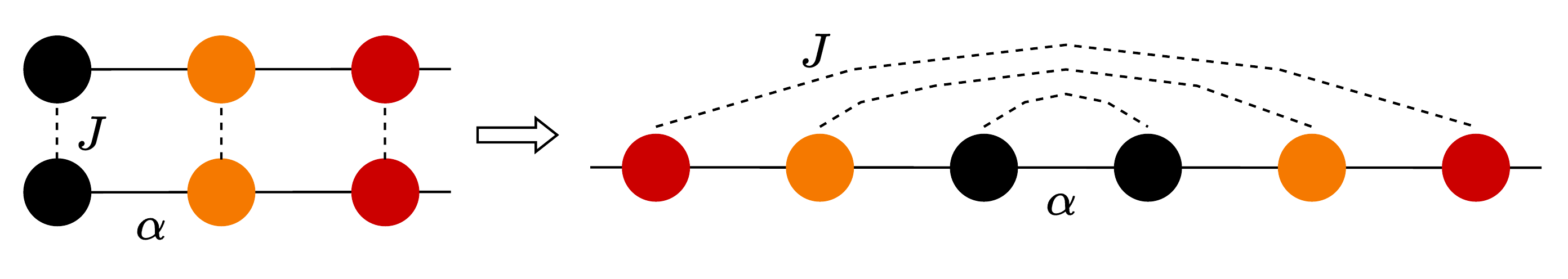}
      \vspace{-0.5em}
      \caption[]
			{\justifying  The long-range entanglement model is depicted. It is given by a qubit ladder mapped to an one dimensional system. Direct neighbors are coupled with the coupling constant $\alpha$, the long-range interactions by $J$.}
      \label{f:longrangemodel}
    \end{figure}

We investigate the influence of non-product initial states and of short- and long-range ansatz schemes on the accuracy of VQE. For this aim, we study the Heisenberg chain with additional long-range couplings, described by the Hamiltonian
    \begin{equation}\label{eq:sl}
        H_{\mathrm{SL}} = \alpha \sum_{i=1}^{n} \left( \vec{S}_i \cdot \vec{S}_{i+1} \right) + J  \sum\limits_{i=1}^{n/2} \left(\vec{S}_i \cdot \vec{S}_{n-(i-1)} \right),
    \end{equation}
    with $n$ being an even total number of qubits in the chain.
    Note that  Eq.\@~\eqref{eq:sl} can be described by a qubit ladder mapped to a one dimensional chain.
    For different values of $\alpha$ and $J$, we obtain a critical to dislocal entanglement  crossover, i.e., the entanglement entropy of a central bipartition of the system behaves as $S(\rho_I) \propto C \log(n)$ for the case $\alpha \gg J$ while $S(\rho_I) \propto n$ for $J \gg \alpha$. The model is visualized in 
    Fig.\@ \ref{f:longrangemodel}.

    \subsection{Random quantum critical point model}
    The long-range entanglement model is inspired by random quantum critical qubit chains where entangled singlets are formed across the whole lattice in an emergent manner \cite{PhysRevLett.43.1434,PhysRevB.22.1305,refae04}.
    We investigate the random Heisenberg chain defined by
    \begin{equation}
        H_{\mathrm{RH}} = \sum_{i=1}^{n} J_i \vec{S}_i \cdot \vec{S}_{i+1} \, .
    \end{equation}
    The couplings $J_i > 0$ are drawn from any non-singular distribution. The randomness
    in the couplings drives the system to a random critical point at long 
    distances \cite{refae04}. 
    Non-local singlets occur in these systems according to the renormalization-group
    (RG) flow: Iteratively, the strongest bond of the system is 
    found and a singlet between the two qubits coupled by this bond is built. 
    The neighboring qubits of the singlet are then coupled by a smaller
    renormalized coupling strength based on second-order perturbation theory.
    Iterating these two steps yields a very good approximation to the ground state of the system with singlets connecting arbitrarily distant sites, which is asymptotically correct for large disorder and large distances.

    %==============================================================================
    % VQE/ Methods
    %==============================================================================

    \section{Methods}
    \label{s:methods}

    \subsection{VQE}
    The quantum circuit of the VQE depending on a set of parameters 
    represents a unitary $U (\vec{\theta})$. 

    The goal of the VQE is to find a set of parameters $\vec{\theta}$, such that $U (\vec{\theta})$ acting on the product state $|0 \rangle ^{\otimes n}$ approximates the ground state of a given Hamiltonian $H$ of a system with $n$ qubits.

    For this, we use quantum-classical optimization loops:
    By applying the circuit with any choice of parameters $\vec{\theta}$ to the initial state
    we obtain a trial wave function 
    $| \psi (\vec{\theta}) \rangle = U (\vec{\theta}) |0 \rangle ^{\otimes n}$.
    With this trial wave function, we can calculate the cost function in a next step.
    The cost function to find the ground state of the system is given by the expectation value of the energy depending on the given wave function 
    $E_{\vec{\theta}} = \langle \psi (\vec{\theta}) | H | \psi (\vec{\theta}) \rangle$.
    The cost function is minimized with a classical optimization algorithm to find the best parameters $\vec{\theta}_{\mathrm{opt}} = \mathrm{argmin}_{\vec{\theta}} \, E_{\vec{\theta}}$.
    With the optimized parameters we then find the best approximation of the ground state of the Hamiltonian 
    $| \psi (\vec{\theta}_{\mathrm{opt}}) \rangle = U (\vec{\theta}_{\mathrm{opt}}) |0 \rangle ^{\otimes n}$
    obtainable with the given quantum circuit.
    
    \subsection{Circuit for local impurity model}
Naturally, the achievable accuracy of the VQE depends on the chosen layout of the quantum circuit.

\begin{figure}[!ht]
      \centering
      \includegraphics[height=4.2cm]{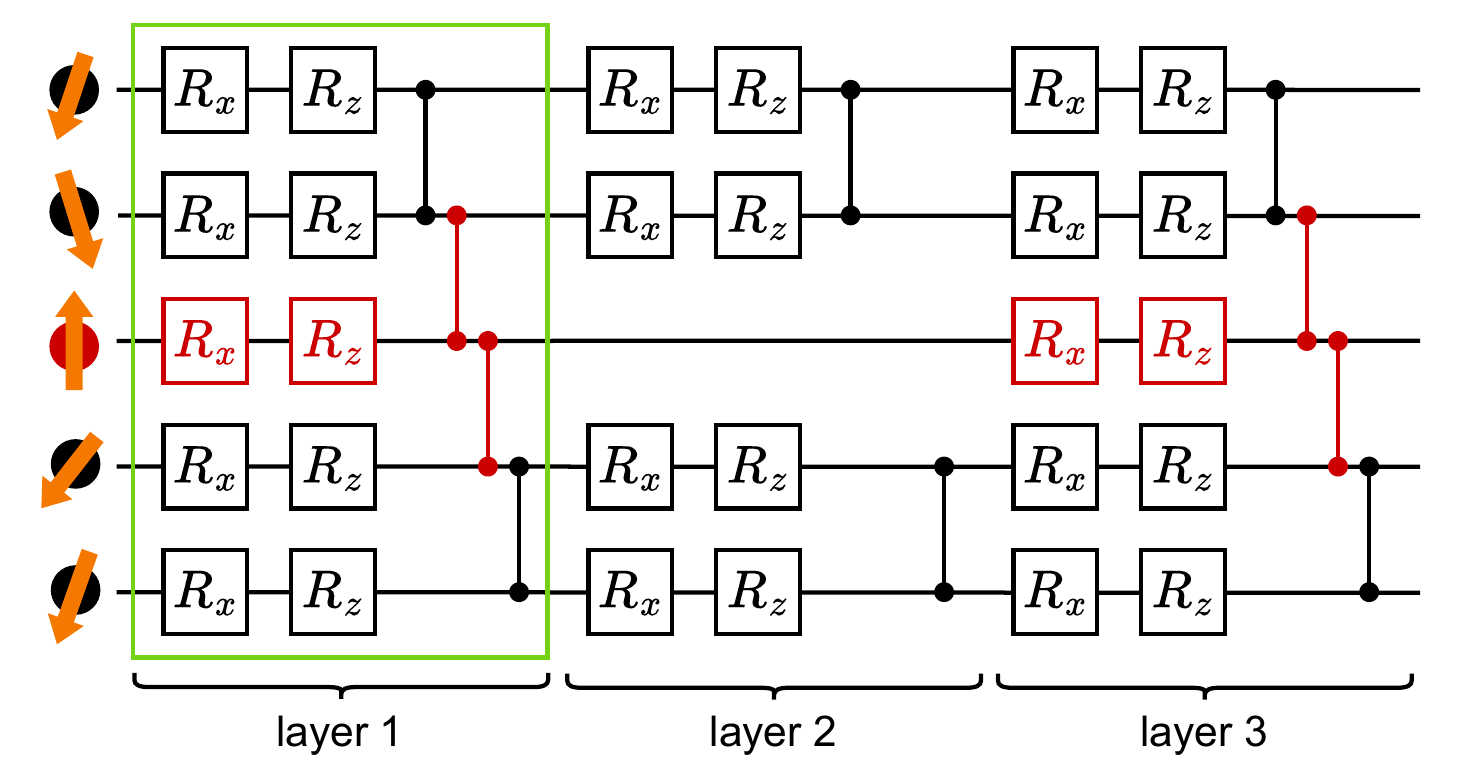}
      \vspace{-0.5em}
      \caption[]
			{\justifying Variational quantum ansatz used in the simulations of the impurity model depicted exemplarily for five qubits and three layers.
      One layer is indicated by the green box consisting of a $R_x$ and a $R_z$ rotation applied to each qubit 
      followed by $R_{zz}$ gates.
      The gates marked in red are only included in the layer if its number is included in $x$, i.e., here $x = [1,3]$.  }
      \label{f:ansatz}
    \end{figure}

    For the local impurity model we will use a quantum circuit built from single-qubit rotations containing $R_x$ and $R_z$ gates on all qubits followed by two-qubit $R_{zz}$ gates applied in a linear chain as shown in Fig. \ref{f:ansatz}.
    This forms one layer of the quantum circuit which can be applied several times.
    We choose the $R_{zz}$ gate instead of the often used CNOT gate, so that a single layer can always be initialized as the identity and $m+1$ layers have at least the same accuracy as $m$ layers.

    The accuracy of the VQE approach for a fixed circuit geometry was studied for these models without impurities in Ref.\@ \cite{bravo20}. Here, we want to focus on the influence of different numbers and positions of entangling gates on the accuracy of the optimized circuits.  

    The most interesting choice is to investigate gates on the central qubit, i.e., we will vary the number of layers in which gates are applied to the impurity qubit.
    As notation we will use $x = $ ["layers including gates applied to the central qubit"].
    The ansatz of the circuit together with this notation is visualized in Fig.\@ \ref{f:ansatz}.

    \subsection{Circuit for long-range entanglement model}
    \label{s:methodsapproach2}

\begin{figure}[htb]
      \centering
      \includegraphics[height=5.0cm]{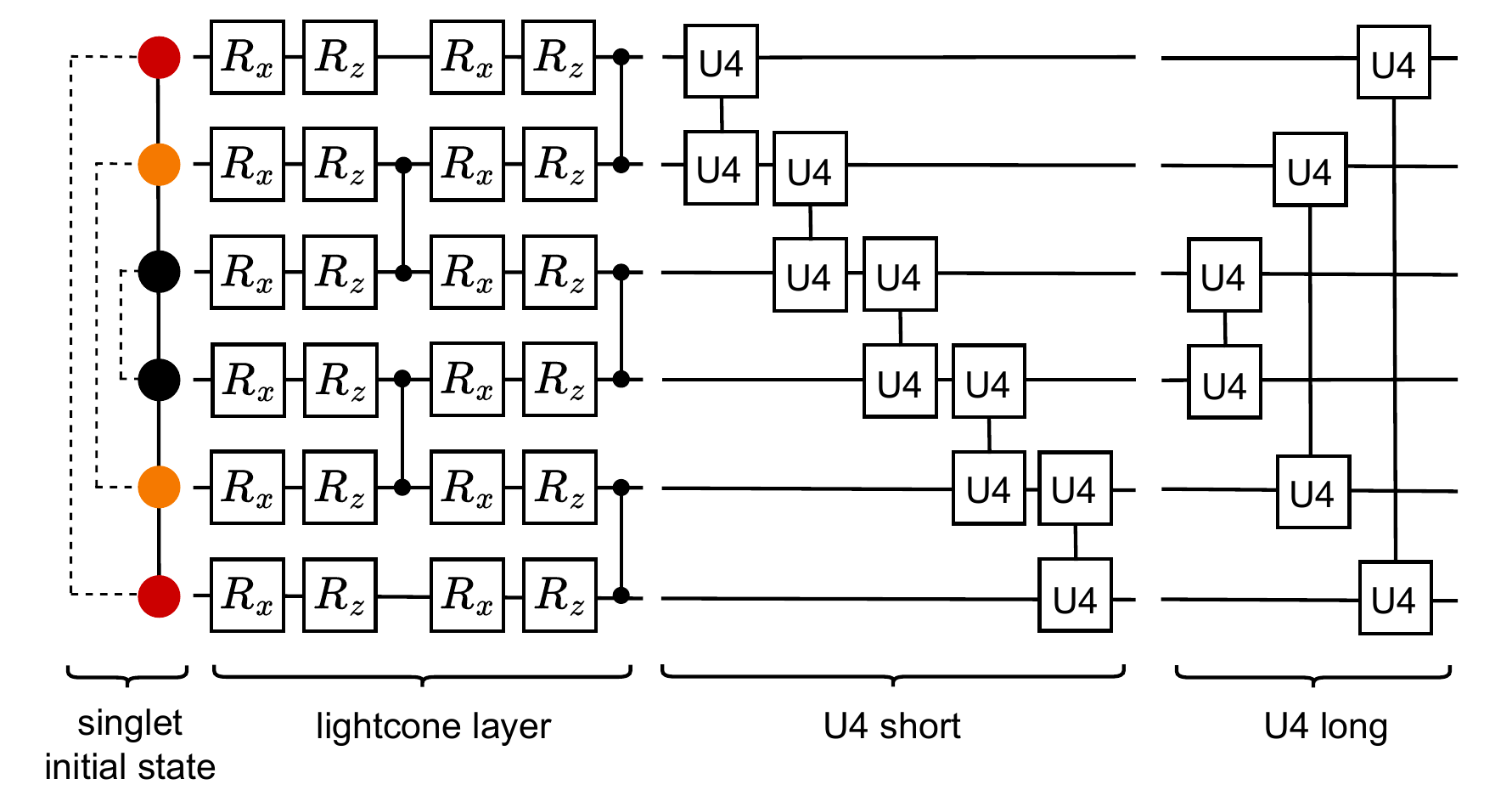}
      \vspace{-0.5em}
      \caption[]{\justifying Variational quantum ansätze used in the simulations of the long-range entanglement model exemplarily depicted
      for 6 qubits. From left to right: singlet initial state is constructed by building singlets between qubits with the
      same color, the dashed line indicates $J$ couplings and the solid line $\alpha$ couplings of Eq.\@~\eqref{eq:sl}; one single lightcone layer consisting of the single qubit rotation gates $R_x$ and $R_z$ and the entangling
      gates $R_{zz}$; set of $U4$ gates entangling neighboring qubits; $U4$ gates entangling qubit pairs which form singlets in the singlet initial state. }
      \label{f:ansatzspinladder}
    \end{figure}

    For the model with long-range entanglement, we investigate the influence of changing the initial state to which the variational ansatz is applied, and the influence of including the long-range entanglement structure of the system into the variational circuit.
    The chosen circuits and gate sets are depicted in Fig.\@~\ref{f:ansatzspinladder}. 
    For the initial state, singlets are built between qubits which 
    match the long-range interactions induced by the second part of Eq.\@~\eqref{eq:sl}.
    This is visualized in Fig.\@ \ref{f:ansatzspinladder} by dots that have the same color.
    We combine this singlet initial state with a varying number of lightcone layers (also visualized in Fig.\@ \ref{f:ansatzspinladder}) and compare it with a lightcone layer applied to a product initial state. The two-qubit gates in the lightcone layer are $R_{zz}$ gates. Additionally, we investigate the effect of matching the short-  and long-range interactions in the variational circuit using a general $U4$ two-qubit gate. The $U4$ gate is implemented optimally as described in Fig.~7 of Ref.~\cite{vatan04}.

    \subsection{Circuit for the random quantum critical point model}
    
\begin{figure}[htb]
      \centering
      \includegraphics[height=5.0cm]{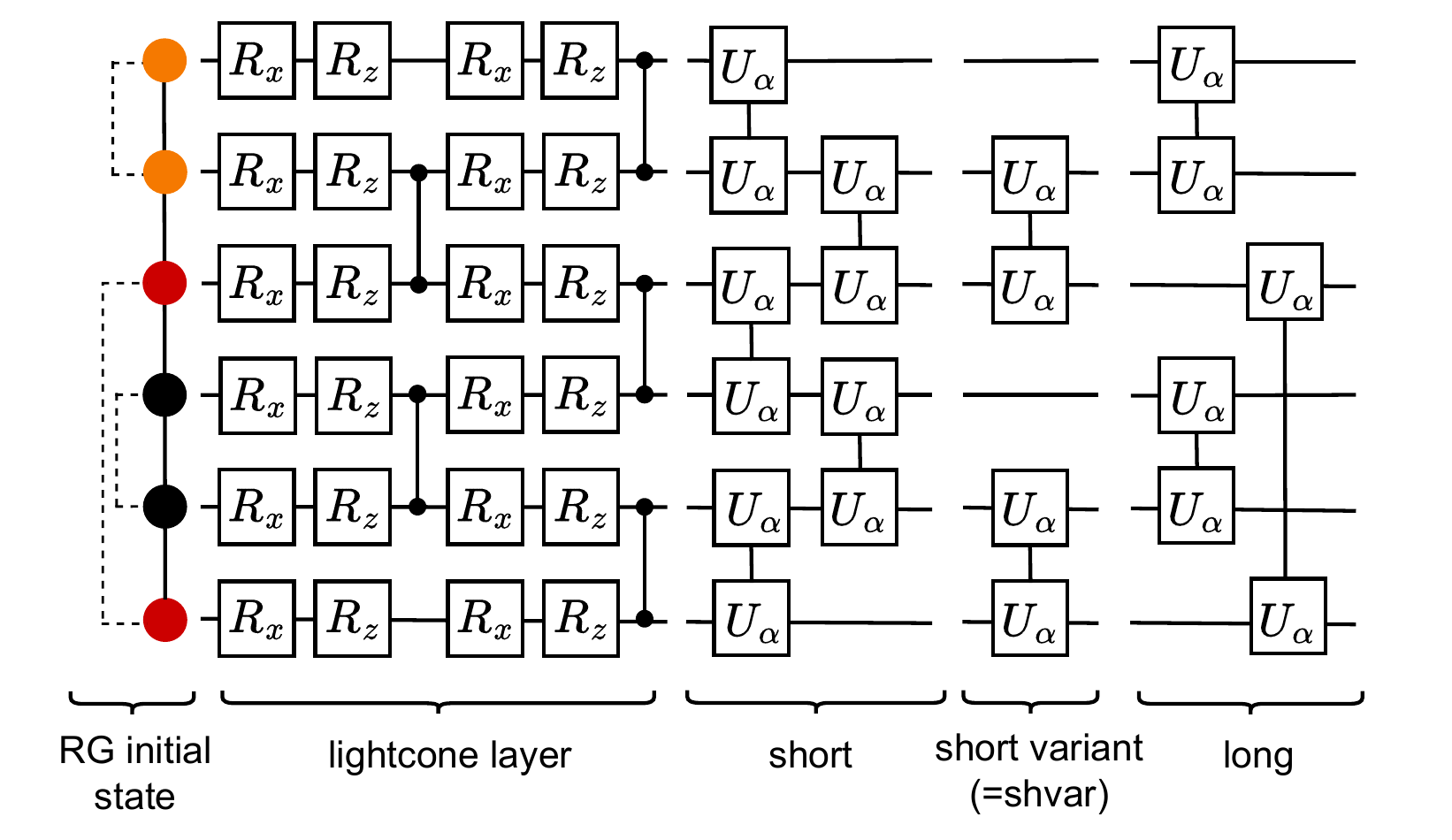}
      \vspace{-0.5em}
      \caption[]{\justifying  Variational quantum ansätze used in the simulations of the random quantum critical point model exemplarily depicted
      for 6 qubits and one possible outcome of the RG flow. From left to right: singlet initial state is constructed by building singlets between qubits with the
      same color, the dashed line indicates the singlets formed and the solid line the $J$ couplings; one single lightcone layer consisting of the single qubit rotation gates $R_x$ and $R_z$ and the entangling
      gates $R_{zz}$; set of $U_\alpha := R_{xx} R_{yy} R_{zz}$ gates entangling qubits with the same structure as in the lightcone layer; $U_\alpha$ gates entangling qubit pairs with strongest couplings except for the ones coupled according to the RG flow - the number of gates is equal to half of the number of RG singlet pairs minus one; $U_\alpha$ gates entangling qubit pairs which form singlets according to the RG flow. }
      \label{f:ansatzdischain}
    \end{figure}

    For the random quantum critical point model, a similar approach to the one of the long-range entanglement
    model is used. The singlets of the singlet initial state are formed according to the long-range interactions resulting from the RG flow. This is exemplarily shown for one configuration in Fig.\@ \ref{f:ansatzdischain}. We also apply lightcone layers to the different initial states and compare the 
    accuracy of the results. 
    To investigate the effect of using the RG singlet structure in the variational part of the circuit, we use
    $U_\alpha := R_{xx} R_{yy} R_{zz}$ gates, which stays within the $S=1$ symmetry sector \cite{Crognaletti_2025}. In the long gate set, the gate is applied to the qubits that form singlets according to the RG flow. The short gate set is structured as in a lightcone layer. A variant of this is to only apply gates to neighboring qubit pairs with the strongest couplings excluding those already coupled within the long gate set. This is done up to a number of half of the RG singlet pairs minus one. In this way combining the short variant with the long gate set gives as many gates as in the short ansatz making a direct comparison possible.
    
    \subsection{Simulation}
    We use the Julia package Yao for quantum circuit simulations \cite{luo20}
and the Broyden–Fletcher–Goldfarb–Shanno (BFGS) method implemented in Optim.jl \cite{Optim} for the classical optimization step. As initial values for the parameters $\vec{\theta}$ for $m+1$ layers, we use the result of $m$ layers and add random numbers close to zero for the additional parameters for the added layer.
    This reduces the total number of required iterations and increases the precision of the results \cite{lyu20}, which is one of the main 
    difficulties here, especially for large numbers of layers.
    The long-range interaction models require several runs with different random numbers 
    to ensure that the results converge.

    For comparison, we compute the exact energy $E$ and exact ground 
    state $| \psi \rangle$ using the eigsolve function from KrylovKit.jl \cite{KrylovKit}, which finds the smallest eigenvalues and their corresponding eigenvectors of a matrix using the Lanzcos algorithm.

    For system sizes of up to 12 qubits, the calculations are performed with double precision making it in principle possible to achieve relative accuracies of up to $10^{-16}$.
    As the number of qubits and layers increases, the simulation time increases rapidly. Therefore, for more than 12 qubits, we simulate the quantum circuit calculations on graphical processors using single precision, which reduces the possible accuracy of the overall calculation to roughly $10^{-8}$.

    %==============================================================================
    % Results Model for Circuit cutting
    %==============================================================================
    \section{Results for local impurity model}
   \label{s:impurity}
    \subsection{Energy}
    \label{s:energy}

    In the following, the optimized energy $E' = E_{\vec{\theta}_{\mathrm{opt}}}$ is calculated depending on, i) total number of layers, ii) number of lattice sites, iii) field strength $h_0^z$ and iv) layers in $x$. 
    The results are compared with the exact result $E$ by computing the relative error $\varepsilon_{\mathrm{rel}} = \left|E -  E' \right| / \left|E\right|$.

\begin{figure}[htb]
      \centering
      \begin{subfigure}[t]{0.48\textwidth}
          \includegraphics[width=8.5cm]{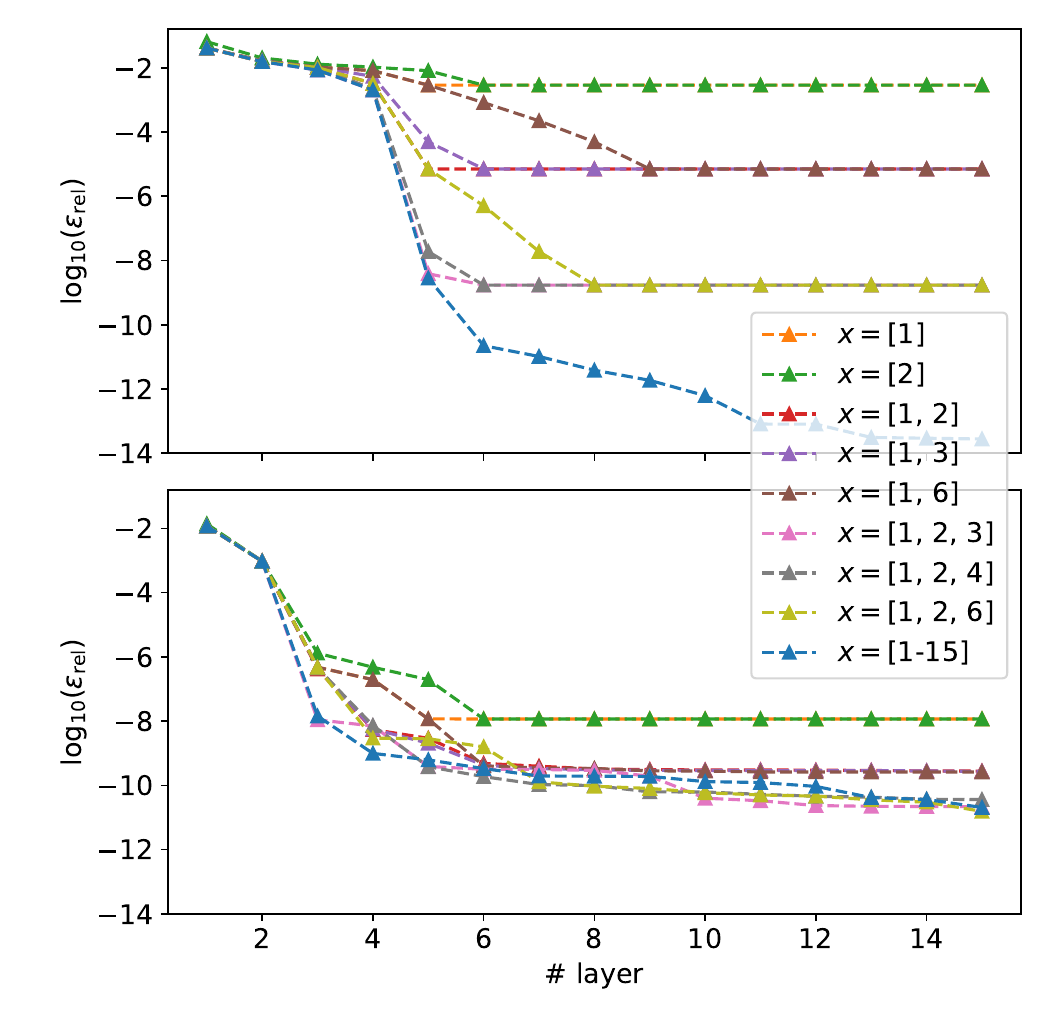}
          \vspace{-0.5em}
          \caption[]{}
          \label{f:buildingplateaustfim}
     \end{subfigure}
     \begin{subfigure}[t]{0.48\textwidth}
        \includegraphics[width=8.5cm]{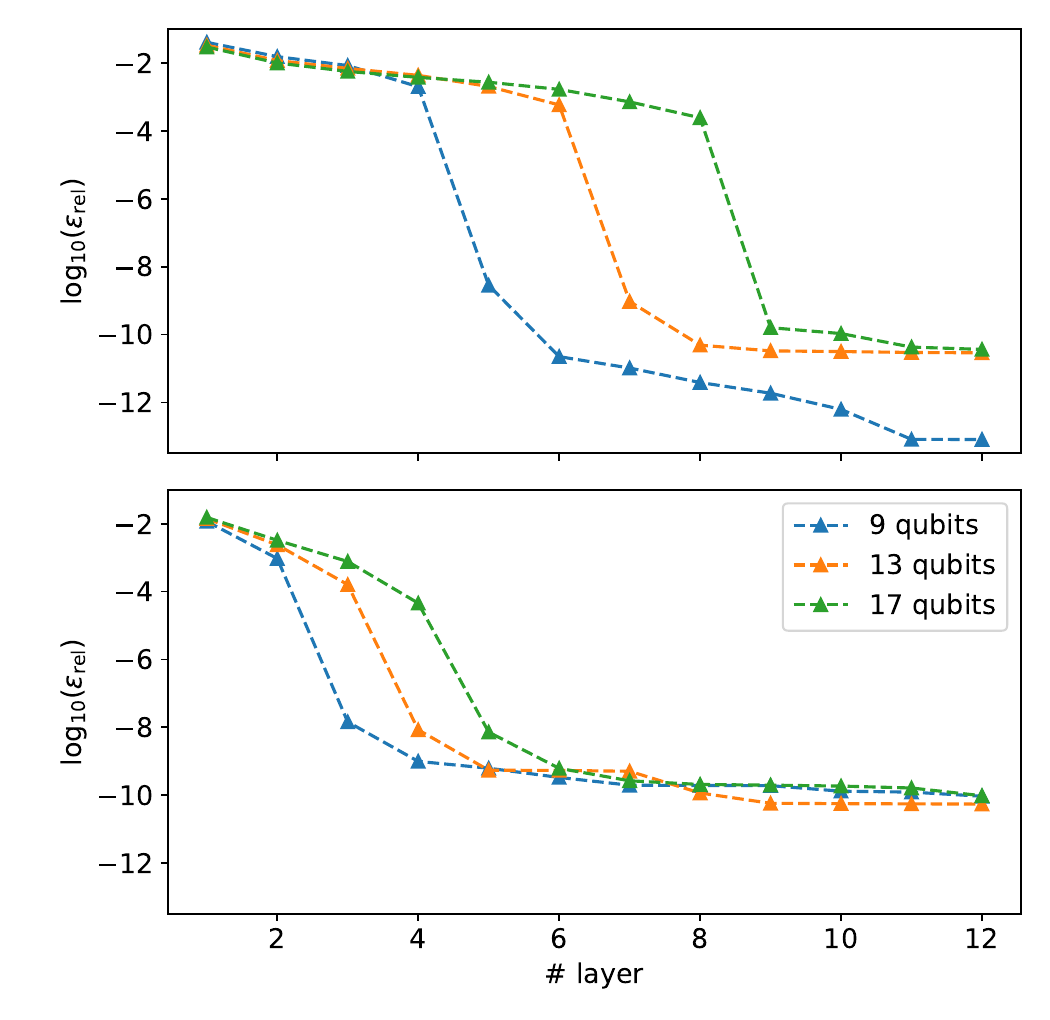}
      \vspace{-0.5em}
      \caption[]{}
      \label{f:regimehalfedtfim}
     \end{subfigure}
     \caption{\justifying 
        (a) Accuracy obtained by the VQE for the TFIM  at $h^x = -1$ without impurity, $h_0^z = 0$, (top) 
              and with impurity, $h_0^z=-10$, (bottom) for 9 qubits
              compared to the exact result in a logarithmic plot. Results for various positions of entangling gates $x$ are shown.
        (b) Accuracy depending on the number of layers for the TFIM with $x = [\mathrm{all}]$ for 
          varying system size without impurity, $h_0^z = 0$, (top) and with impurity, $h_0^z=-10$, (bottom).
     }
    \end{figure}

    An overview of the results for $9$ qubits in the TFIM model is shown in Fig.\ \ref{f:buildingplateaustfim}.
    Accuracy plateaus which form after a sufficient number of layers are our primary finding. The value of the plateau only depends on the absolute number of entangling layers: By adding a layer to $x$ the value of the plateau is decreased, i.e., the accuracy of the ansatz improves.  However, the earlier the entanglement gates are applied in the algorithm, the sooner the plateau value is reached.

    Comparing  the impurity problem and the problem
    without impurity in Fig.\@ \ref{f:buildingplateaustfim}, one can see that the value of the plateau improves drastically when including an impurity even for the VQE with only a single entangling layer. Here, we find an increase of the groundstate energy accuracy of several orders of magnitude.

    In critical phases, two regimes have been observed for the convergence behavior \cite{bravo20}: For small numbers of 
    layers the accuracy improves slowly independent of the system size. This regime is called finite-depth regime. After a critical number of layers, a finite-size regime starts. In this 
    regime, the accuracy improves again exponentially.
    This behavior can also be seen in our results in Fig.\@~\ref{f:buildingplateaustfim} for 9 qubits.
    Comparing the results with and without impurity, a major difference is recognized with respect to
    the transition to an exponential convergence in accuracy: 
    With impurity the number of layers needed to reach exponential improvement is halved. To investigate this further, we computed the accuracy depending on the number of layers for $x = [\mathrm{all}]$, where 'all' means that every layer is included in $x$, for various system sizes, visualized in Fig.\@ \ref{f:regimehalfedtfim}.

    \begin{figure}[htb]
      \centering
      \begin{subfigure}[t]{0.48\textwidth}
          \includegraphics[width=8.3cm]{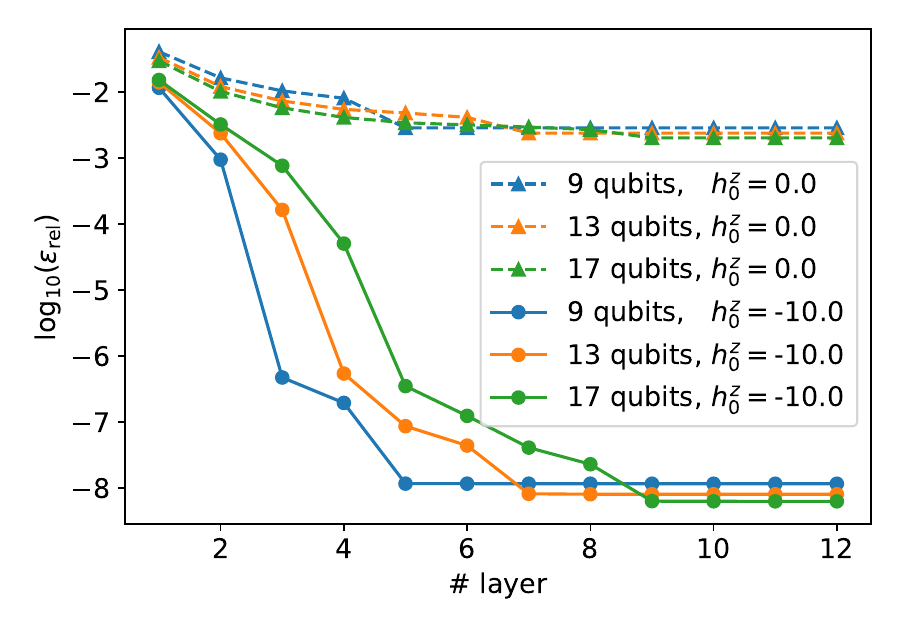}
          \vspace{-0.3em}
          \caption[]{ }
          \label{f:nlayertoplateau}
      \end{subfigure}
      \begin{subfigure}[t]{0.48\textwidth}
          \includegraphics[width=8.7cm]{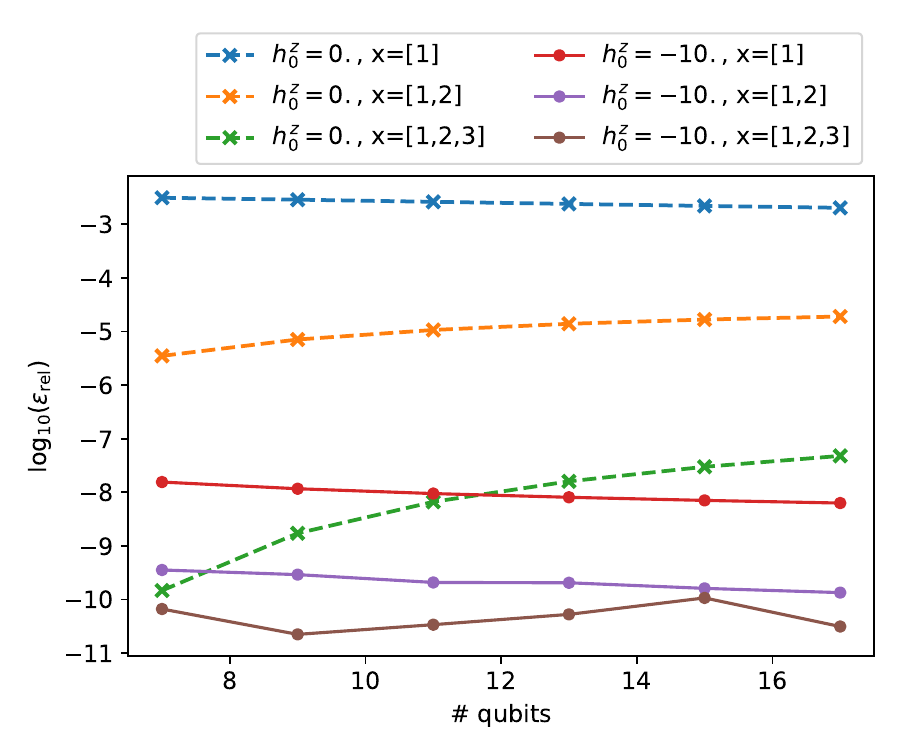}
          \vspace{-1.5em}
          \caption[]{}
          \label{f:readoffaccuracytfim}
      \end{subfigure}
      \caption{\justifying
      (a) Accuracy depending on the number of layers for the TFIM with $x = [1]$ for varying system size without impurity, $h_0^z = 0$, and with impurity, $h_0^z=-10$.
          By including the impurity, the finite-depth regime is halved but the number of layers to 
          reach the plateau stays the same.
      (b) Logarithmic accuracy obtained by the VQE compared to the exact result for different system sizes as well as
          number of layers in $x$ for the TFIM at the critical point. The number of entries in $x$ indicates the number of cuts 
          necessary to divide the circuit. Thus, one can read off the number of cuts necessary to achieve a given accuracy. 
      }
    \end{figure}

    Without impurity, we can reproduce the results from Ref.\@ \cite{bravo20}:
    First, we see a finite-depth regime followed by a finite-size regime depending linearly on 
    the system size, where an exponential improvement in accuracy is achieved (the plateaus 
    at high numbers of layers are formed due to finite precision). 
    The separation into these two regimes is an important difference compared to gapped phases.
    With impurity, we can see in Fig.\@~\ref{f:regimehalfedtfim} also for other system sizes that the finite-depth regime is halved.
    This behavior can be explained by the bisection of the chain induced by the 
    impurity.
    The variational algorithm needs only half of the layers as without impurity to realize that the system
    is finite.

    In Fig.\@ \ref{f:nlayertoplateau}, we investigate the case $x =[1]$ for various system sizes. Without impurity, we can reach only a limited accuracy due to the high plateau value. In contrast, the plateau with impurity is much lower and is almost independent of system size. Interestingly, the finite-size regime, i.e., the exponential behavior, is reached earlier with impurity, but the plateau requires the same number of layers to be reached as without impurity.

    To scale this approach, it is important to investigate the plateau values for different $x$ for increasing system sizes to reach a better insight into how much the plateau values improve with impurity. In Fig.\@ \ref{f:readoffaccuracytfim}, this is shown for
    $x=[1], x=[1,2]$ and $x=[1,2,3]$.
    We can see that by increasing the number of layers in $x$ the accuracy of the result increases for all chain lengths.
    Without impurity, the plateau values are much higher than with impurity. All results seem to slowly converge to finite values. This indicates that both, impurity strength $h_0^z$ and number of layers between subsystems $x$ remain relevant on all length scales.

    %==============================================================================
    % Results Model for Circuit cutting
    %==============================================================================
    \subsection{Entanglement entropy and spectrum}
    \label{s:spectrum}

\begin{figure}[htb]
      \centering
      \includegraphics[width=8.5cm]{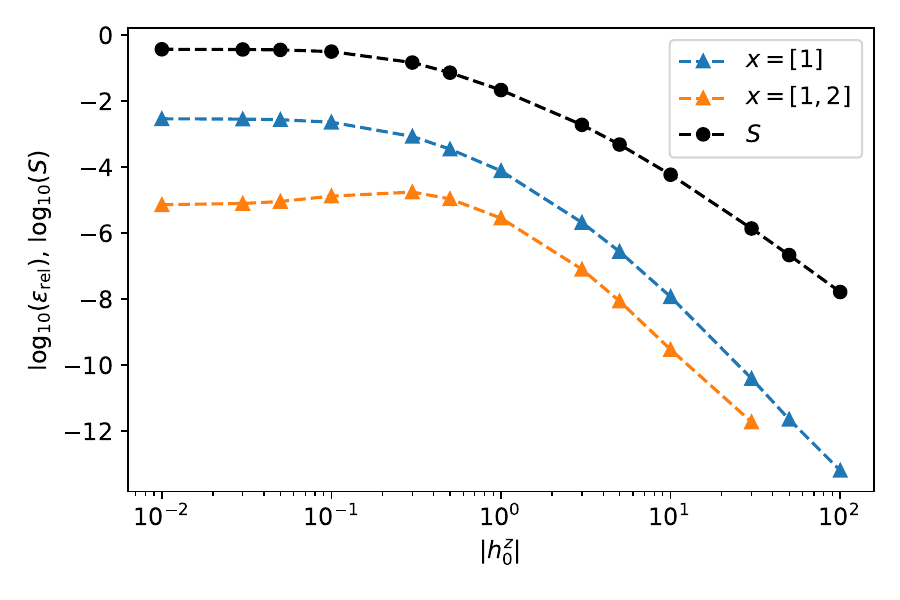}
      \vspace{-1.5em}
      \caption[]{\justifying Logarithmic accuracy obtained with the VQE at the plateau for $x=[1]$ and $x=[1,2]$ compared to the entanglement entropy for varying impurity strength for the TFIM with 9 qubits.}
      \label{f:entangentrtfim}
    \end{figure}

    To further understand the previous results in terms of entanglement properties, we first investigate the entanglement entropy. We, therefore, compute the reduced density matrix of the exact ground state $|\psi \rangle$
    \be 
      \rho_\mathrm{A} = \mathrm{Tr}_\mathrm{B} |\psi \rangle \langle \psi| 
    \ee
    with $A$ being the left half of the chain and $B$ the right half including the central qubit.
    The entanglement entropy is given by
    \be
      S = -\Tr \left(\rho_\mathrm{A} \ln(\rho_\mathrm{A}) \right) \, .
    \ee
    To better understand the improvement in accuracy when adding an impurity, we 
    plot the exact entanglement entropy as function of $h_0^z$ and the relative energy error $\varepsilon_{\mathrm{rel}}$ for $x=[1]$ and $x=[1,2]$ in Fig.\@ \ref{f:entangentrtfim}.

    For increasing magnetic field strength, the entanglement entropy
    decreases and at the same time the accuracy improves, following the same trend. Interestingly, increasing the number of layers in $x$ seems to have an almost constant effect on the relative accuracy improvement for weak as well as for strong impurity field strength $h_0^z$. 

    In a second step, we investigate the entanglement spectrum to 
    better understand the improvement in accuracy with increasing number of layers in $x$.
    The entanglement spectrum is given by the eigenvalues of the entanglement 
    Hamiltonian
    \be 
      H_{\mathrm{ent}} = - \ln (\rho_\mathrm{A}) \, .
    \ee

    We compare the entanglement spectrum of the exact ground state $|\psi \rangle$ with the entanglement spectrum of the variational state $| \psi (\vec{\theta}) \rangle$.

\begin{figure}[htb]
      \centering
      \begin{subfigure}[c]{0.48\textwidth}
          \includegraphics[width=8.5cm]{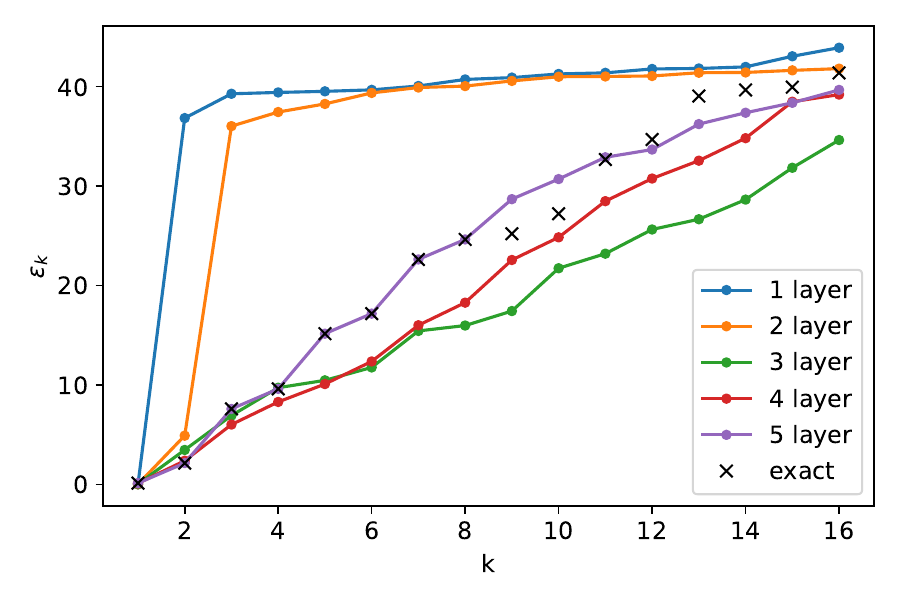}
          \vspace{-0.5em}
          \caption[]{ }
          \label{f:esnlayervar}
      \end{subfigure}
      \begin{subfigure}[c]{0.48\textwidth}
          \includegraphics[width=8.5cm]{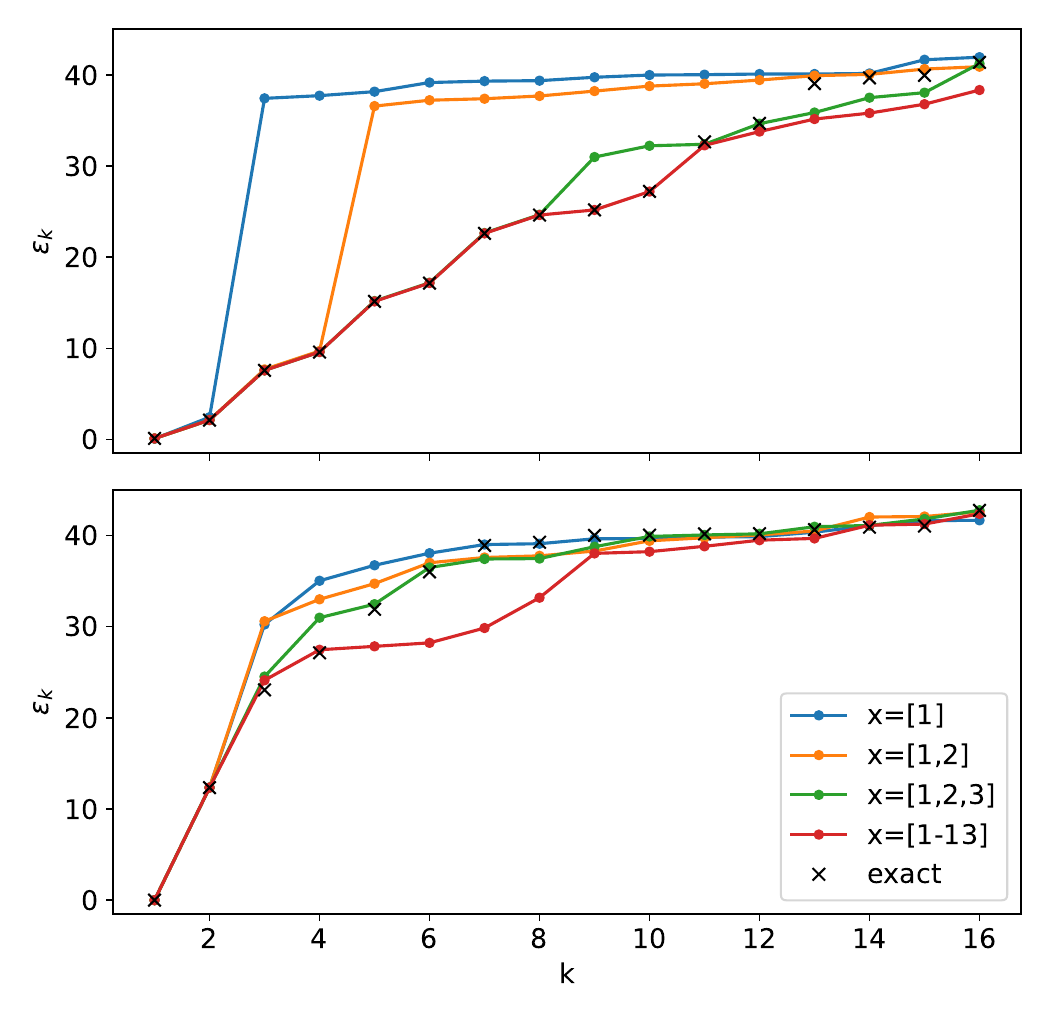}
          \vspace{-1.0em}
          \caption[]{}
          \label{f:esxvartfim}
      \end{subfigure}
      \caption{\justifying
      (a) Eigenvalues $\varepsilon_k$ of the entanglement Hamiltonian for the TFIM at the critical
          point with 9 qubits
          without impurity for $x= [\mathrm{all}]$ depending on the number of layers compared to the exact 
          solution.
      (b) Eigenvalues $\varepsilon_k$ of the entanglement Hamiltonian for the TFIM at $h^x = -1$ with 9 qubits
          without, $h_0^z = 0$, (top) and with impurity, $h_0^z=-10$, (bottom) impurity for 13 layer for different $x$ compared to the exact 
          solution.
      }
    \end{figure}

   In Fig.\@ \ref{f:esnlayervar}, we first study the entanglement spectrum for the TFIM with 9 qubits without impurity for $x=[\mathrm{all}]$ depending on the total number of layers. 
    It can be seen that for small numbers of layers only the first eigenvalue is captured well and up to four layers the accuracy improves slowly.
    With five layers, however, all eigenvalues up to a value of around 25 are calculated correctly by the VQE calculation.
    This fits the transition of the regimes, as for 9 qubits without impurity the 
    exponential behavior is reached after the fourth layer, see Fig.\@~\ref{f:regimehalfedtfim}.
    Larger eigenvalues are not reproduced correctly even with five layers due to the 
    given precision in the calculation, as high eigenvalues of the energy entanglement
    spectrum correspond to extremely small values in the reduced density matrix.
    Hence, only very precise numerics is able to evaluate them reliably.
    
    In Fig.\@ \ref{f:esxvartfim}, the entanglement spectrum is shown for the TFIM with 9 qubits
    with and without impurity for 13 layers for different $x$.
    By increasing the number of layers included in $x$, we iteratively capture more eigenvalues with the VQE approach starting with the  small, most significant eigenvalues.
    In particular, in the case without impurity the number of correctly found eigenvalues is doubled with each additional layer in $x$.
    The largest eigenvalues are not calculated correctly even in the case of $x=[\mathrm{all}]$. This is again due to the precision of the calculation.

    When comparing the exact eigenvalues of the systems with and without impurity, it becomes clear that the TFIM without impurity has many more eigenvalues with a significant contribution to the spectrum. Therefore, any VQE algorithm has to reproduce these eigenvalues correctly for an accurate description of the groundstate, while in the case with impurity only the first few values are necessary. Therefore, even for the strongly simplified circuits with only one or two entangling layers, the algorithm can be expected to yield accurate groundstate energies although the approach captures only the first few eigenvalues of the entanglement spectrum. The other eigenvalues are much larger, hence much less significant.

    In addition to the aforementioned results, we also investigated the behavior of energy accuracy and of entanglement entropy and spectrum for the XXZ chain. The findings are comparable to those observed for the TFIM, with only minor deviations. The results are presented in Appendix \ref{a:xxzxhain}.

    \subsection{Magnitude of gradient}
    
    So far, we showed that in impurity models we can significantly reduce the number of entangling gates
    at this impurity while maintaining high accuracy. This is advantageous for the use of circuit 
    cutting techniques and for reducing potentially mitigating errors introduced by gate operations.
    Here, we show that the reduced ansatz additionally addresses the problem of barren plateaus. 
    To this end, we examine the gradient for both the full and the reduced ansatz for different numbers of 
    layers and qubits for the TFIM model with impurity.
    
\begin{figure}[htb]
      \centering
      \begin{subfigure}[t]{0.48\textwidth}
          \includegraphics[width=8.5cm]{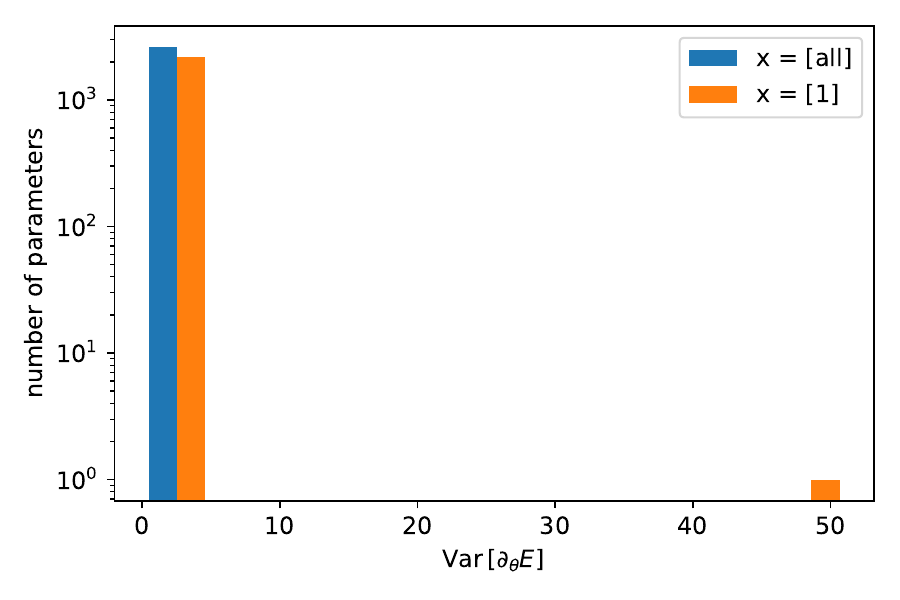}
          \vspace{-0.5em} 
          \caption[]{}
          \label{f:grad1}
      \end{subfigure}
      \begin{subfigure}[t]{0.48\textwidth}
          \includegraphics[width=8.5cm]{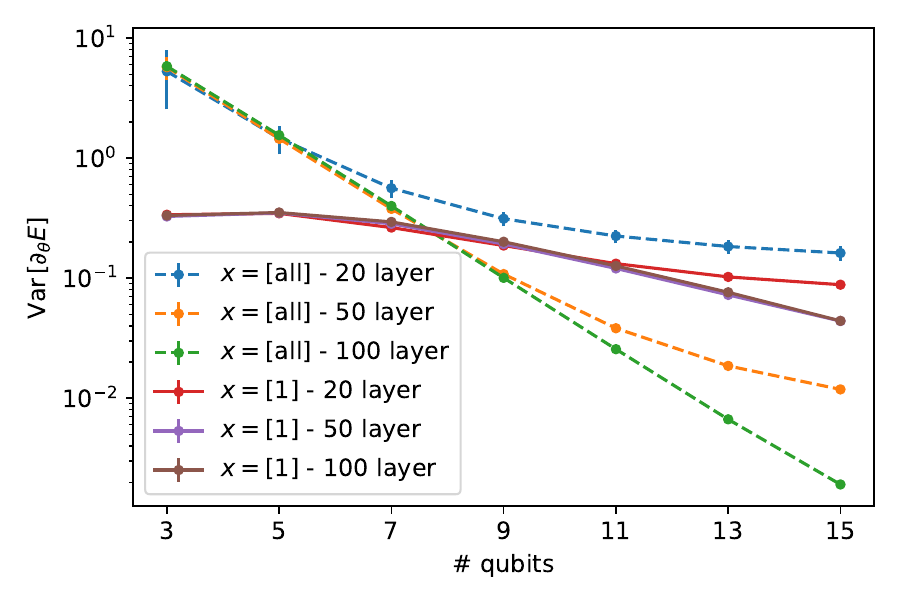}
          \vspace{-0.5em}
          \caption[]{}
          \label{f:grad2} 
      \end{subfigure}
      \caption{\justifying
      (a) Number of parameters with similar variance of gradient for $x=[1]$ (reduced ansatz) and $x=\,$[all] (full ansatz) for 9 qubits and 200 layers in the case with impurity. The large variance value corresponds to the first rotation parameter at the impurity qubit.
      (b) Average variance of gradient over all parameter choices excluding the parameter of the first impurity rotation angle. The error bars are the variance across the different gradient components.
      }
    \end{figure}

    While the mean of the gradient for random choices of $\vec{\theta}$ is zero, the variance can be larger. 
    A larger variance is advantageous for the minimization as non-zero elements are needed to find a direction for the next minimization step. 
    It was shown that the variance approaches zero for increased circuit depth and system
    size \cite{McClean2018,PRXQuantum.3.010313}.
    We first investigate the distribution of the variance of the gradient over all parameter choices in Fig.\@ 
    \ref{f:grad1} for one choice of circuit depth and system size.
    In the case of the reduced ansatz, there is one parameter which has a much larger variance than all other parameters. This is the first rotation angle applied to the impurity and this large value 
    is also present for other system sizes and circuit depths.
    This means that the minimization algorithm can optimize in that direction.
    All other parameters have much smaller variances, being all in the same order of magnitude.
    Therefore, we want to have a closer look at the behavior of those parameters to get an idea about the 
    risk of barren plateaus for the case that the 
    parameter with large variance is already optimized.

    An examination of the variance of the gradient of the remaining parameters is shown in Fig.\@ \ref{f:grad2} with respect to system size. It reveals that the slope of the reduced ansatz is, on average, reduced in comparison to that of the full ansatz.
    This means that with the reduced ansatz the problem of barren plateaus is decreased even without the parameter with large variance as there are statistically more non-zero values for minimization.
    Therefore, we can conclude that our reduced ansatz scheme is to be preferred over the full ansatz in terms
    of circuit cutting techniques, reduced errors and improved optimization capability.

    %==============================================================================
    % Results Advantages of singlet state
    %==============================================================================
    \section{Results for the long-range entanglement model}
    \label{s:singletstate}

    \subsection{Energy}

\begin{figure}[htb]
      \centering
      \begin{subfigure}[t]{0.48\textwidth}
          \includegraphics[width=8.5cm]{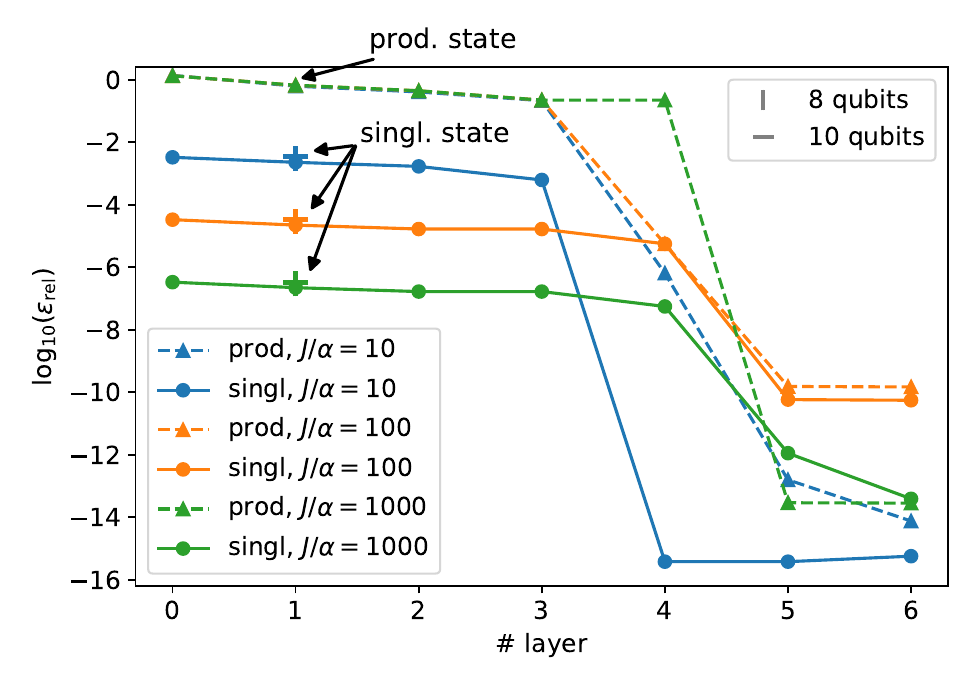}
          \vspace{-0.5em}
          \caption[]{ }
          \label{f:alphavar}
      \end{subfigure}
      \begin{subfigure}[t]{0.48\textwidth}
          \includegraphics[width=8.5cm]{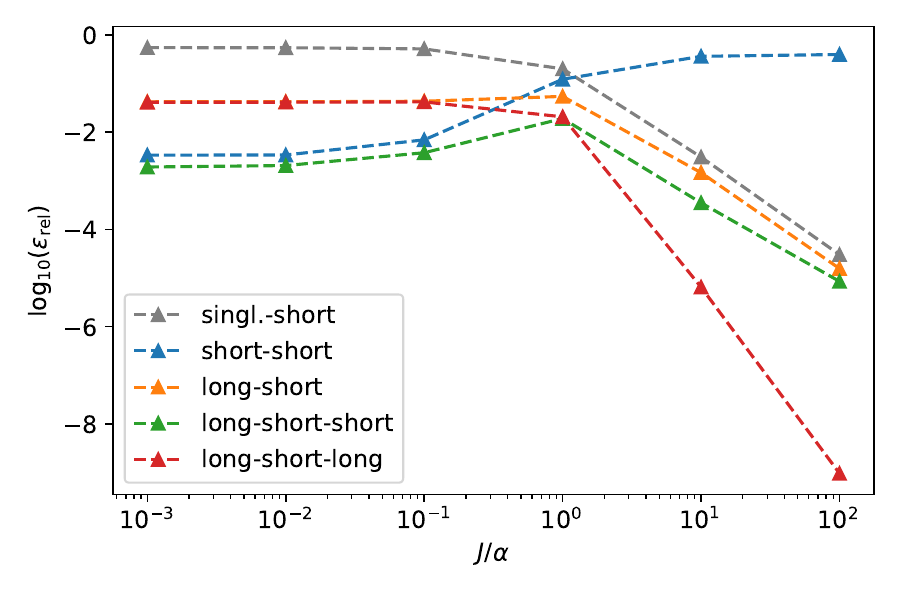}
          \vspace{-0.5em}
          \caption[]{ }
          \label{f:differentansatze}
      \end{subfigure}
      \caption{\justifying
      (a) Accuracy of energy obtained with VQE for 6 qubits for various ratios $J / \alpha$. Additionally, for one layer the results for 8 and 10 qubits are depicted.
      (b) Comparison of accuracies obtained with different combinations of $U4$ short and $U4$ long gate sets (see Fig.\@ \ref{f:ansatzspinladder}) forming one layer for varying
          ratio of $\alpha$ and $J$ for $10$ qubits. Additionally, the results with initial singlet state and one $U4$ short gate set (singl.-short) are shown in grey for comparison.
      }
    \end{figure}

    We start to investigate the model in Eq.\@ \eqref{eq:sl} for large ratios $J/\alpha \gg 1$. As in the previous section, we focus on the relative energy error. The results are presented in Fig.\@ \ref{f:alphavar} for 6 qubits where we compare an initial product state with an initial singlet state. Different numbers of lightcone layers are added to the initial states. We choose to switch to the lightcone layer because the convergence was improved sligthly compared to the ansatz structure used for the impurity model.

    We can observe that increasing the value of $J/\alpha$ leads to an improvement in the results with singlet initial states.
    This difference is directly present without a variational layer. 
    Additionally, the findings for 8 and for 10 qubits are plotted for one layer with singlet initial state. 
    We can see that there is no dependence on the system size, as the accuracy is maintained for different qubit chain lengths when comparing the same value of $\alpha$.
    This is consistent with the results of Ref.\@ \cite{sun23}, which demonstrated for correlated topological phases that with an initial state fitting the given structure of the system one variational layer is sufficient to achieve
    good results independent of the system size.
    If the desired accuracy is in the range of $10^{-2}$ to $10^{-6}$, the solution with singlet 
    initial state and one variational layer is already good enough. In contrast, starting from a product state, multiple variational 
    layers are necessary to achieve the same level of accuracy.

    However, the accuracy depends directly on the ratio $J/\alpha$.
    For $\alpha \approx J$, the accuracy decreases towards the value obtained with product initial state.
    At the same time, the number of layers needed to converge to higher accuracy does not differ 
    significantly between the two initial states, i.e., in Fig.\@ \ref{f:alphavar} 4 to 5 layers are necessary in both cases.
    For larger system sizes, even more layers are needed.
    It follows that either if (i) $\alpha \approx J$ or if (ii) $\alpha < J$ and higher accuracies are desired, the singlet initial state does no longer yield an improvement to the product initial state.
    It is therefore of interest to identify an ansatz that can be used to achieve higher accuracy with a single layer.
    Moreover, with the ansatz chosen so far consisting of $R_{zz}$-entangling gates, we find bad convergence
    behavior during optimization.
    Using $U4$ instead of $R_{zz}$ gates showed to improve the convergence.
    Consequently, we employ a $U4$-layer ansatz to find the best solution with only one layer.
    Note that the $U4$ gate has many more free parameters leading to an improved accuracy but also a longer
    runtime.

    A first approach is to replace the initial singlet state by a dimerized variational layer.
    We study if having optimizable parameters between the long-range coupled qubits improves the results further.
    We use $U4$-gates between the qubits coupled by $J$. 
    This then builds one $U4$ long gate set as described in Sec.\@ \ref{s:methodsapproach2}.
    In Fig.\@ \ref{f:differentansatze}, the accuracies obtained with different combinations of the $U4$ long gate set with a $U4$ short
    gate set are compared for different ratios of $\alpha$ and $J$. 
    For comparison, also the results with initial singlet state and one short gate set are shown in grey.
     
    For the singlet initial state combined with an $U4$ short gate set (singl.-short), we observe that for $J \gg \alpha$ this choice leads to results already in the order of $10^{-4}$ analogously
    to the results obtained before with a lightcone layer applied to the singlet initial state. For smaller $J$,
    the accuracy decreases.
    Comparing this with the ansätze with product initial state, we see that using the $U4$ long gate set, especially more than once, yields an improvement to
    the singlet initial state for all ratios $J / \alpha$.
    In general, we find three regimes: For $J \ll \alpha$, the $U4$ short gate set gives the best
    results and an additional $U4$ long gate set only has a small impact. 
    For $J \approx \alpha$, we have an intermediate regime and for $J \gg \alpha$, the ansatz with two $U4$ long gate sets gives a large
    improvement compared to other ansätze. For $J / \alpha = 100$, we even reach accuracies in the order of $10^{-8}$ with only a single layer.
    In a next step, we check if this result is also valid for larger system sizes.

\begin{figure}[htb]
      \centering
      \begin{subfigure}[t]{0.48\textwidth}
          \includegraphics[width=8.5cm]{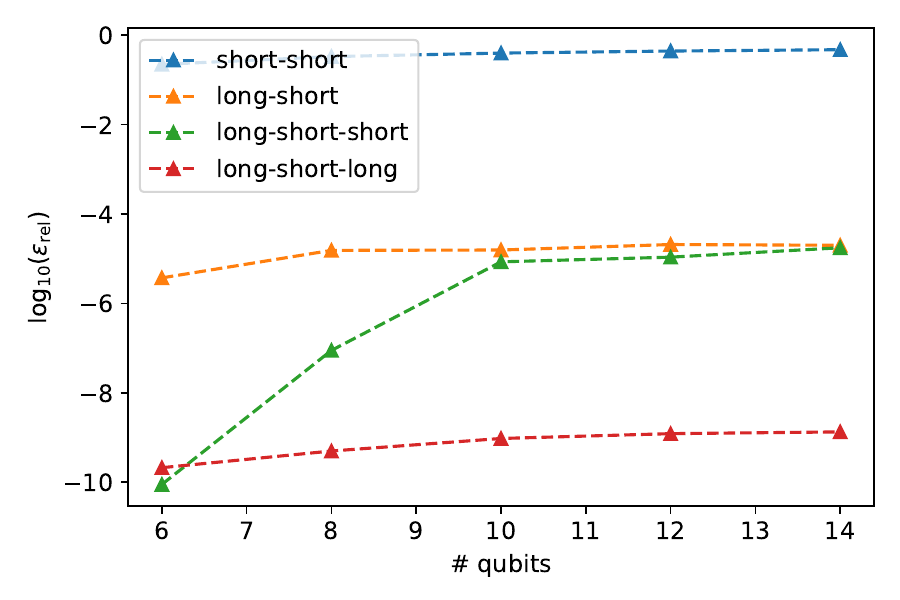}
          \vspace{-0.5em}
          \caption[]{}
          \label{f:J100_differentsizes}
      \end{subfigure}
      \begin{subfigure}[t]{0.48\textwidth}
          \includegraphics[width=8.5cm]{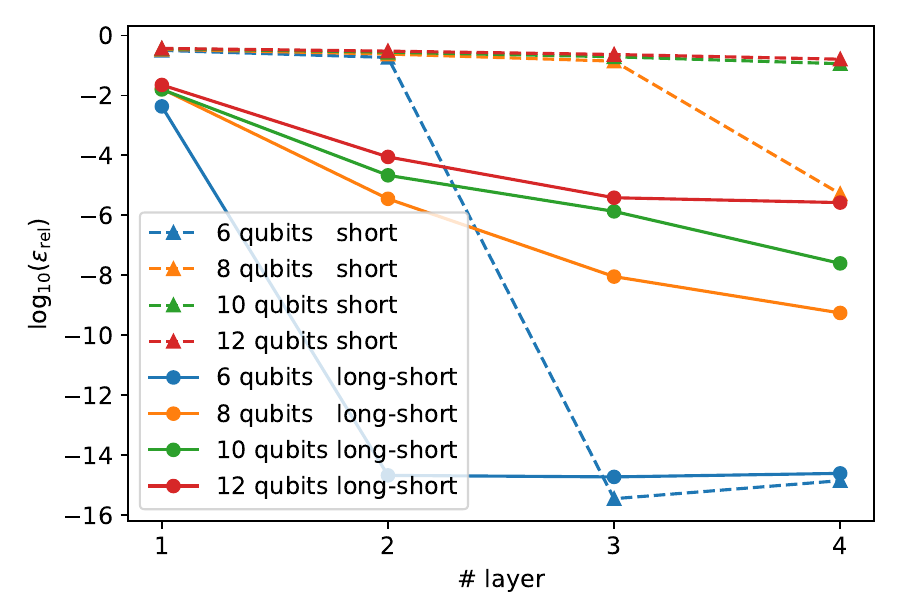}
          \vspace{-0.5em}
          \caption[]{ }
          \label{f:J3_differentlayers}
      \end{subfigure}
      \caption{\justifying
      (a) System size dependence of the different ansätze consisting of different combinations of $U4$ short and $U4$ long gate sets (see Fig.\@ \ref{f:ansatzspinladder}) for $J / \alpha = 100$.
      (b) Accuracy depending on the number of layers for a $U4$ short ansatz and a  $U4$ long-short ansatz (see Fig.\@ \ref{f:ansatzspinladder}) for $J / \alpha = 3$ for various 
          system sizes.
      }
    \end{figure}

    In Fig.\@ \ref{f:J100_differentsizes}, we show the dependence of the different ansätze on the system size
    for $J / \alpha = 100$.
    The accuracy obtained with the long-short-short ansatz decreases for larger system sizes. 
    This indicates that for small systems the short gate set is still sufficient to capture the long-range interactions
    which is no longer the case for larger system sizes.
    However, the other ansätze converge to a constant accuracy for larger system sizes.
    As for the solution with singlet initial state and one variational layer shown in Fig.\@ \ref{f:alphavar}, we find a system size independent
    behavior.

    It follows that we can already identify the long-short-long ansatz as a good choice for $J \gg \alpha$.
    However, we can see in Fig.\@ \ref{f:differentansatze} that for $J \rightarrow \alpha$ the accuracy 
    is worse than $10^{-4}$ even with the long-short-long ansatz.
    Therefore, we investigate in a next step whether applying a small number of layers of an ansatz including $U4$ long gate sets 
    provides an improvement compared to using only a single layer, which was not the case for the solutions 
    obtained with an ansatz consisting only of gates between neighboring qubits as visible in Fig.\@ \ref{f:alphavar}.
    We compare the solution of an $U4$ short ansatz with the solution of an long-short ansatz for up to 4 layers.
    The results are depicted in Fig.\@ \ref{f:J3_differentlayers} for $J / \alpha = 3$.

    Comparing the results with the short ansatz, we find the typical behavior of a critical system with 
    a finite-depth and a finite-size regime. That implies that the larger the system size is, the more layers
    are required to achieve an improvement in accuracy.
    For the long-short layer, we observe an exponential improvement in the finite-depth regime in contrast to the 
    nearly constant behavior in the case of the short ansatz.
    Consequently, much better results can be obtained with only a few layers, which constitutes a great advantage 
    especially for large system sizes.
    The number of layers necessary to obtain a certain accuracy depends on the ratio
    $J/\alpha$: The larger the ratio, the fewer layers are required.

    In conclusion, the singlet initial state provides an initial improvement to the accuracy. 
    However, a combination of variational long-range and short-range gates facilitates convergence for the whole phase space. Therefore, for achieving good accuracies with few parameters/ layers, including the 
    long-range structure into the variational circuit is the best choice.

    \subsection{Entanglement spectrum}

\begin{figure}[htb]
      \centering
      \begin{subfigure}[c]{0.48\textwidth}
          \includegraphics[width=8.5cm]{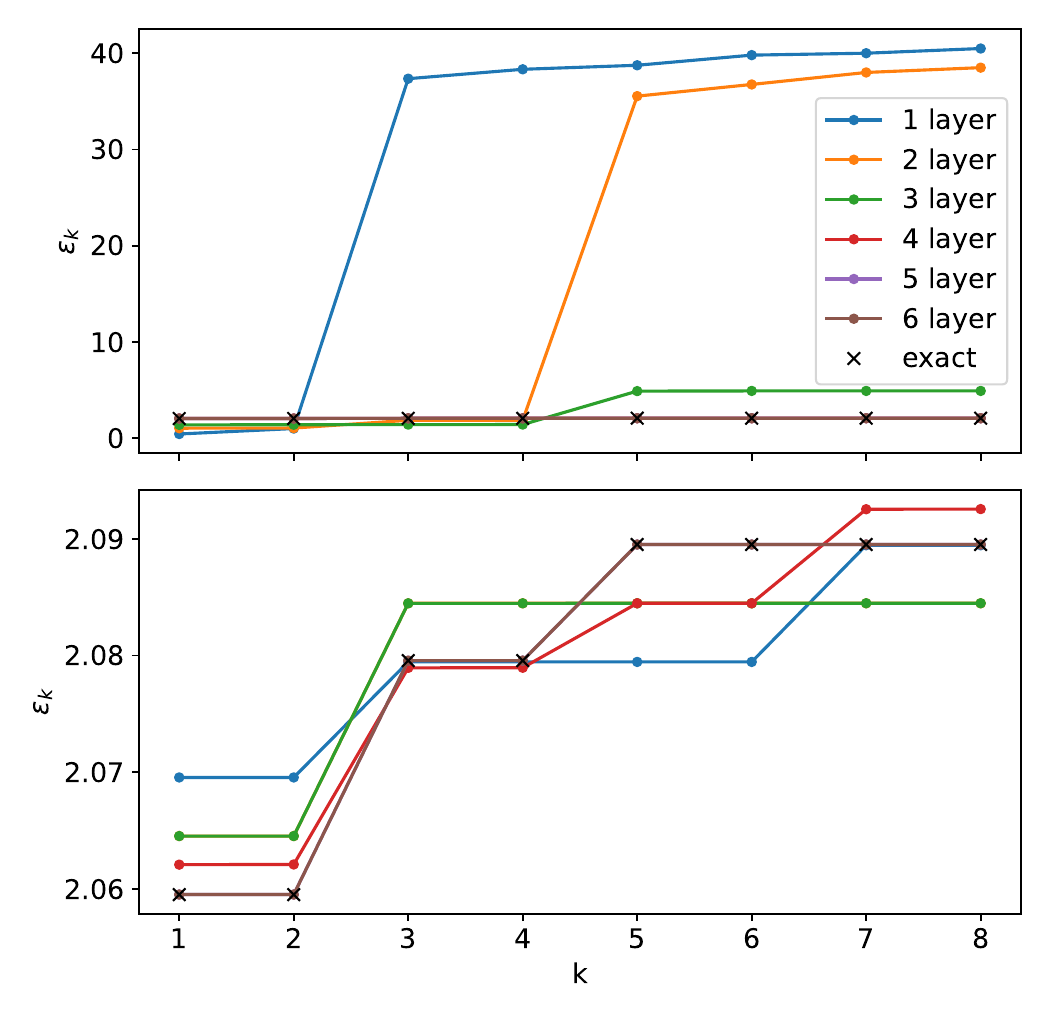}
          \vspace{-1.5em}
          \caption[]{}
          \label{f:entangspecsmallalpha}
      \end{subfigure}
      \begin{subfigure}[c]{0.48\textwidth}
          \includegraphics[width=8.5cm]{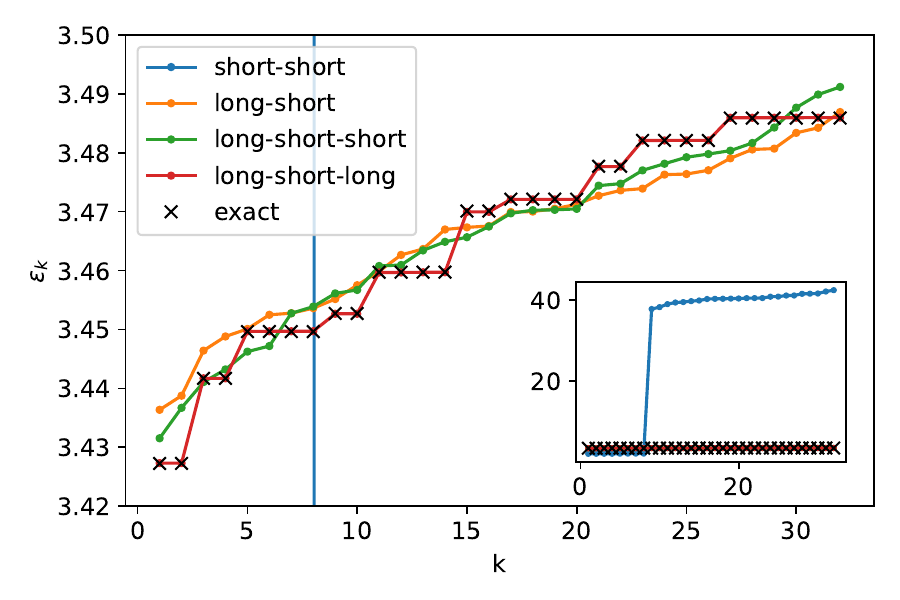}
          \vspace{-1.5em}
          \caption[]{}
          \label{f:entangdifferentansatze}
      \end{subfigure}
      \caption{\justifying
      (a) Entanglement spectrum for $J / \alpha = 100$ and 6 qubits. The upper plot shows the results 
          with product initial state, the lower plot the results with singlet initial state.
          Note the different intervals of the $\epsilon_k$-axis.
      (b) Entanglement spectrum for the different ansätze for 10 qubits and $J = 100$. The inset shows the behavior of the short-short ansatz.
      }
    \end{figure}

    As before, we also investigate the entanglement spectrum to obtain a better understanding of the different convergence behaviors.
    First, the entanglement spectrum for $J / \alpha = 100$ of the data shown
    in Fig.\@ \ref{f:alphavar} is plotted for 6 qubits in Fig.\@ \ref{f:entangspecsmallalpha}.
    
    First of all, when comparing the exact spectrum with the one shown in Fig.\@~\ref{f:esxvartfim} we observe that the range of the values is small compared to the ones 
    from the impurity model.
    This can be attributed to the high entanglement inherent in the qubit ladder model.
    In general, a global convergence is observed instead of finding the lowest eigenvalues first, which is due to the long-range entanglement.
    With singlet initial state the eigenvalues found with VQE are close to the exact ones 
    already with one layer.
    In contrast, with product initial state the eigenvalues are close to the exact ones from the fourth 
    layer onwards.
    This fits to the results of the accuracy in Fig.\@ \ref{f:alphavar}.

    Next, in Fig.\@ \ref{f:entangdifferentansatze} we compare the accuracy of the calculated 
    entanglement spectra of the different ansätze in Fig.\@ \ref{f:differentansatze} for $J/\alpha = 100$.
    The results obtained with the short-short ansatz are not even close to the exact ones,
    which changes as soon as one $U4$ long gate set is included in the ansatz. For the long-short-long ansatz, all eigenvalues are
    found correctly, while the number of parameters for this approach is approximately the same as for the short-short approach.
    It can thus be inferred that the long-short-long approach is the most suitable.

    %==============================================================================
    % Random quantum critical point model
    %==============================================================================
    \section{Random quantum critical point model}
    \label{s:random_model}

    A more difficult model than treated in the last section is the random quantum critical point model as singlets 
    are formed randomly depending on the couplings.

    We consider always several configurations for the random couplings and 
    take the average value for all quantities of interest of the different runs.

    The random coupling values $J_i$ are sampled from the distribution
    \begin{equation}
        P(J) = \frac{1}{\delta} J^{-1+\delta^{-1}} 
    \end{equation}
    as suggested in Ref.\@ \cite{laflo05} with a disorder strength $\delta \geq 1$. The larger $\delta$, the 
    larger is the disorder of the system.

    We start again by inspecting the result for a prepared initial state together with one lightcone
    layer for different values of $\delta$. The chosen initial state consists of singlets between qubit pairs which 
    form a singlet according to the RG flow:
    The two qubits coupled by the largest value $J_i$ build a singlet. Under the condition $J_i \ll J_{i-1}, J_{i+1}$, the neighboring qubits are then coupled by a Heisenberg interaction with strength
    \cite{refae04}
    \begin{equation}
        J'_{i-1, i+2} = \frac{J_{i-1} J_{i+1}}{2 J_i} \, 
    \end{equation}
    resulting from second-order perturbation theory.
    These two steps are iterated until all qubits are paired to singlets.
    
\begin{figure}[htb]
      \centering
      \begin{subfigure}[t]{0.48\textwidth}
          \includegraphics[width=8.5cm]{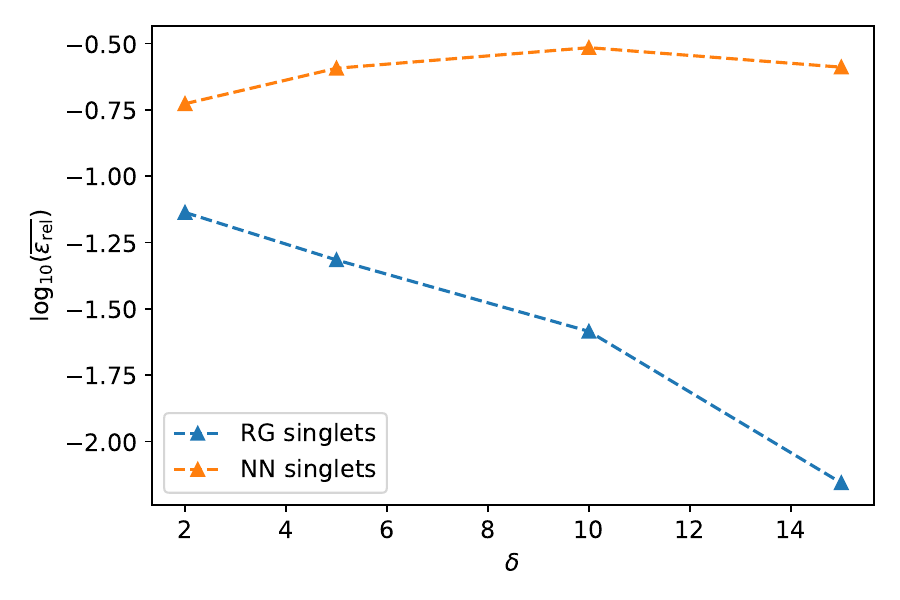}
          \vspace{-0.5em}
          \caption[]{}
          \label{f:dis_delta}
      \end{subfigure}
      \begin{subfigure}[t]{0.48\textwidth}
          \includegraphics[width=8.5cm]{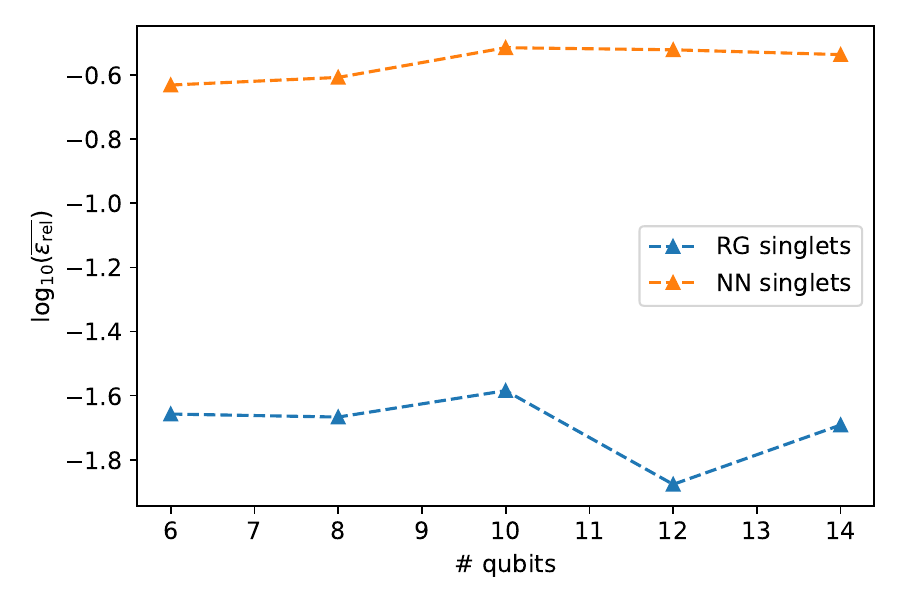}
          \vspace{-0.5em}
          \caption[]{}
          %todo: 14 spins läuft noch durch -> ergänzen
          \label{f:dis_size}
      \end{subfigure}
      \caption{\justifying
      (a) Energy accuracy for one lightcone layer for two initial states depending on the disorder strength $\delta$ for $10$ qubits. It is averaged over $100$ configurations.
      (b) Energy accuracy for two initial states and one lightcone layer for $\delta = 10$ averaged over 100 configurations. Except for statistical fluctuations, we see an independence of the system size.
      }
    \end{figure}

    This is compared with singlets formed between neighboring 
    qubits. The results are displayed in Fig.\@ \ref{f:dis_delta} for a system size of $10$ qubits. The larger
    $\delta$ is, the better is the result for the RG initial state compared to the initial state with nearest neighbor (NN) singlets. 
    This was to be expected, as large values of $\delta$ correspond to a distribution with rather disparate $J$ values, in which case the RG approach works particularly well.
    Additionally, the result matches with the results from the last section, where we saw the same
    for increasing the ratio of $J$ and $\alpha$.

    For a fixed value of $\delta$, the results remain independent of system size, see Fig.\@ \ref{f:dis_size}. 

\begin{figure}[htb]
      \centering
      \begin{subfigure}[t]{0.48\textwidth}
          \includegraphics[width=8.5cm]{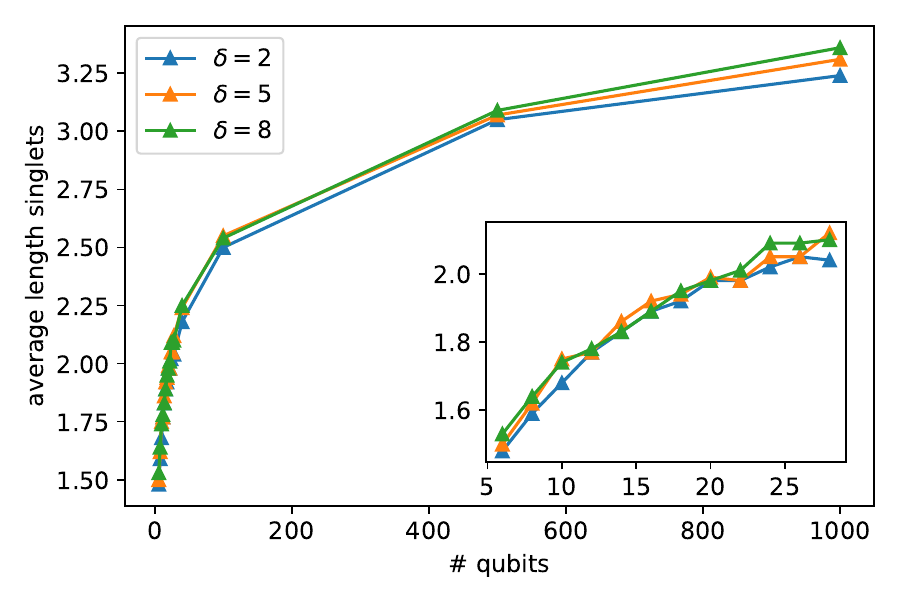}
          \vspace{-0.5em}
          \caption[]{}
          \label{f:dis_singlets}
      \end{subfigure}
      \begin{subfigure}[t]{0.48\textwidth}
           \includegraphics[width=8.5cm]{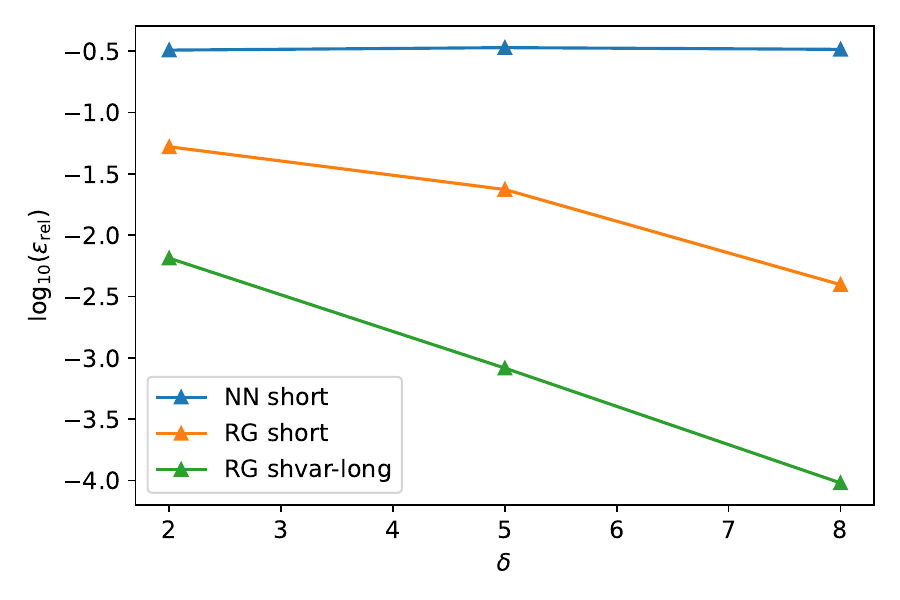}
          \vspace{-0.5em}
          \caption[]{}
          \label{f:tensor_network}
      \end{subfigure}
      \caption{\justifying
      (a) The average length of singlets given by the RG flow is plotted depending on the system size averaged over 1000 configurations for different disorder strengths $\delta$.
      (b) Energy accuracy of the results for using NN singlets in the initial state 
        combined with a short layer, for using RG singlets in the initial state combined with a $U_\alpha$ short layer
        and for
        using RG singlets in the initial state combined with a layer consisting of short variant (shvar) and a $U_\alpha$ long gate set depending on the disorder parameter $\delta$ for $120$ qubits. The different gate sets are depicted in 
        Fig.\@~\ref{f:ansatzdischain}.
      }
    \end{figure}

    When applying more layers to these initial states, starting from the second layer on, the results converge and the chosen initial state does no longer make any difference. 
    %Plot hierzu ergänzen oder nur so in Textform? oder statt letzten beiden Plots nur einen analog zu fig 12 (da dort "alles auf einmal")?
    Therefore, as for the long-range
    entanglement model, a next step to reach an overall improvement would be to include the entanglement 
    structure of the system not only in the initial state, but in the circuit itself.
    However, the average length of the singlets grows only logarithmically with the system size as can be seen 
    in Fig.\@~\ref{f:dis_singlets}. 

    To test whether incorporating the structure into the circuit itself works
    well, we investigate a system size of $120$ qubits. To be able to simulate such a large system, we switched
    to tensor network calculations using quimb implemented in python \cite{quimb}. In Fig.\@ \ref{f:tensor_network}, we compare the accuracy of the results for (i) using NN singlets in the initial state 
    combined with a short layer, for (ii) using RG singlets in the initial state combined with a short layer and for (iii)
    using RG singlets in the initial state combined with a layer consisting of short variant and a long gate set. The gate sets are depicted in Fig.\@ \ref{f:ansatzdischain}. Note that all three ansätze used have the same number of gates and parameters. The accuracy is calculated for three values of $\delta$. 

   The results show that the accuracy using an RG initial state increases with increasing $\delta$ in comparison to using singlets between neighboring qubits as the initial state. This outcome aligns with the result observed for $10$ qubits in Fig.\@~\ref{f:dis_delta}. Most importantly, if the structure resulting from the RG flow is adopted within the variational circuit itself, a subsequent enhancement is observed, exhibiting a similar magnitude to the previous one. It is evident that a well-chosen initial state and variational circuit significantly enhances the performance, given that the number of parameters remains constant across all three variants.

For investigating how this behavior scales with more layers, a system size of about $1000$ qubits is an interesting choice. Since simulations with a classical computer are no longer feasible at this scale, it would be instructive to evaluate it on a real quantum device. We leave this question open for future investigations.

    %==============================================================================
    % Conclusion
    %==============================================================================
    \section{Conclusions}
    \label{s:conclusion}

    In this paper, we provided an extensive study on how to adapt variational circuits 
    to the entanglement structure 
    of impurity models and of models with intrinsic and emergent long-range interactions.
    This is motivated by the assumption that the ground state entanglement implies the convergence of the variational ansatz.
    We investigated both the convergence of the ground state energy and of the entanglement
    spectrum depending on the number of layers and the ansatz choice.
    
    For the impurity models, we found that plateaus are formed in the accuracy of the energy  depending on the number of layers. The value of the plateau
    depends on the absolute number of entangling gates between the subsystems and on the strength of 
    the impurity. This knowledge can be used to avoid unnecessary 
    entangling gates, making it possible to simulate larger systems on noisy quantum devices.
    We showed that our ansatz gives an improvement in terms of circuit cutting
    techniques, reduced errors and improved optimization capability.
    These results are of high interest for various impurity models, e.g., the Anderson model used to 
    describe electrons in (heterostructure) quantum dots \cite{ander61}, 
    the Kondo model which has emerged as an important tool for understanding the behavior of strongly correlated systems and continues to offer considerable promise \cite{kondo64, piqua23}, and for simulations of molecules \cite{bhart22}. 

    We further found for the long-range entanglement model that including both the short-range and the 
    long-range behavior of the model in the variational circuit improves the convergence of the VQE. This implies that fewer layers are required to achieve a desired accuracy and therefore
    improves the circuit in depth.
    Long-range interactions play a role in numerous physical phenomena, making them a crucial field of investigation \cite{defen23, buchh23}. 

    Inspecting the entanglement spectrum calculated with VQE, an iterative convergence is visible for the impurity models meaning that one eigenvalue after the other is captured.
    In comparison, the long-range entanglement model shows global convergence, with 
    the eigenvalues being much closer to each other than in the case of the impurity models.
    These results open a new path in understanding the convergence of variational quantum circuits based on the global properties of the entanglement spectrum.

    Finally, we demonstrated that ideas inspired by renormalization group theory can be used to guide the design of efficient low-depth circuits for tackling the challenging class of random quantum critical points. By incorporating the hierarchical structure of the RG approach, our ansatz schemes significantly outperforms uninformed alternatives, which we showed for simulations on systems as large as $120$ qubits. Thus, physically motivated ansätze beyond the established variational Hamiltonian approach \cite{PhysRevA.92.042303,Park2024hamiltonian} are capable of capturing emergent long-range entanglement and point to the feasibility of implementing such strategies on near-term quantum hardware. 

    Future work may be directed towards the application of the developed ansätze to the 
    aforementioned systems on NISQ devices, as the models are of great physical interest. As our results 
    specifically address the issues associated with NISQ devices, they offer the potential to gain interesting
    insight into the various models using current quantum computing platforms.

    %==============================================================================
    % Acknowledgments 
    %==============================================================================
    \begin{acknowledgments} 
    This work has been supported by the Quantum Fellowship Program of the German Aerospace Center (DLR). We thank F. Eickhoff and K. Lively for helpful discussions.
    \end{acknowledgments}

    \bibliography{liter11.bib}
		
    %==============================================================================
    % Appendix 
    %==============================================================================
    \begin{appendix} 
	  	\section{Second impurity model: XXZ chain}
      \label{a:xxzxhain} 
    As comparison to the TFIM we also study the behavior of the XXZ model with an impurity to show that the results
    obtained are rather generic.
    Its Hamiltonian is given by
    \begin{equation}\label{xxz}
      H_{\mathrm{XXZ}} = \sum_{i=-L/2}^{L/2} \left( \sigma_i^x \sigma_{i+1}^x + \sigma_i^y \sigma_{i+1}^y + \Delta \sigma_i^z \sigma_{i+1}^z \right) + h_0^z \sigma_0^z 
    \end{equation}
    with an anisotropy $\Delta$ and the Pauli matrices $\sigma^{x,y,z}$.
    Similarly to the TFIM, we add a term for a longitudinal field at the central qubit.
    This model is gapped for $|\Delta| \, > 1$.
    For $\Delta < -1$, the qubits are aligned ferromagnetically, while for 
    $\Delta > 1$, they are aligned antiferromagnetically along the $z$-direction.
    For $-1 \leq \Delta \leq 1$, the system is in a critical phase and is 
    described by a conformal field theory with central charge $c=1$.
    We choose $\Delta = 0.5$, which is far off from the Heisenberg point as well as from the gapped phase.

    As before, we compare the results obtained with VQE with results found numerically by diagonalization
    and we use open boundary conditions to separate the chain with increasing impurity strength.

\begin{figure}[htb]
      \centering
      \begin{subfigure}[t]{0.48\textwidth}
          \includegraphics[width=8.5cm]{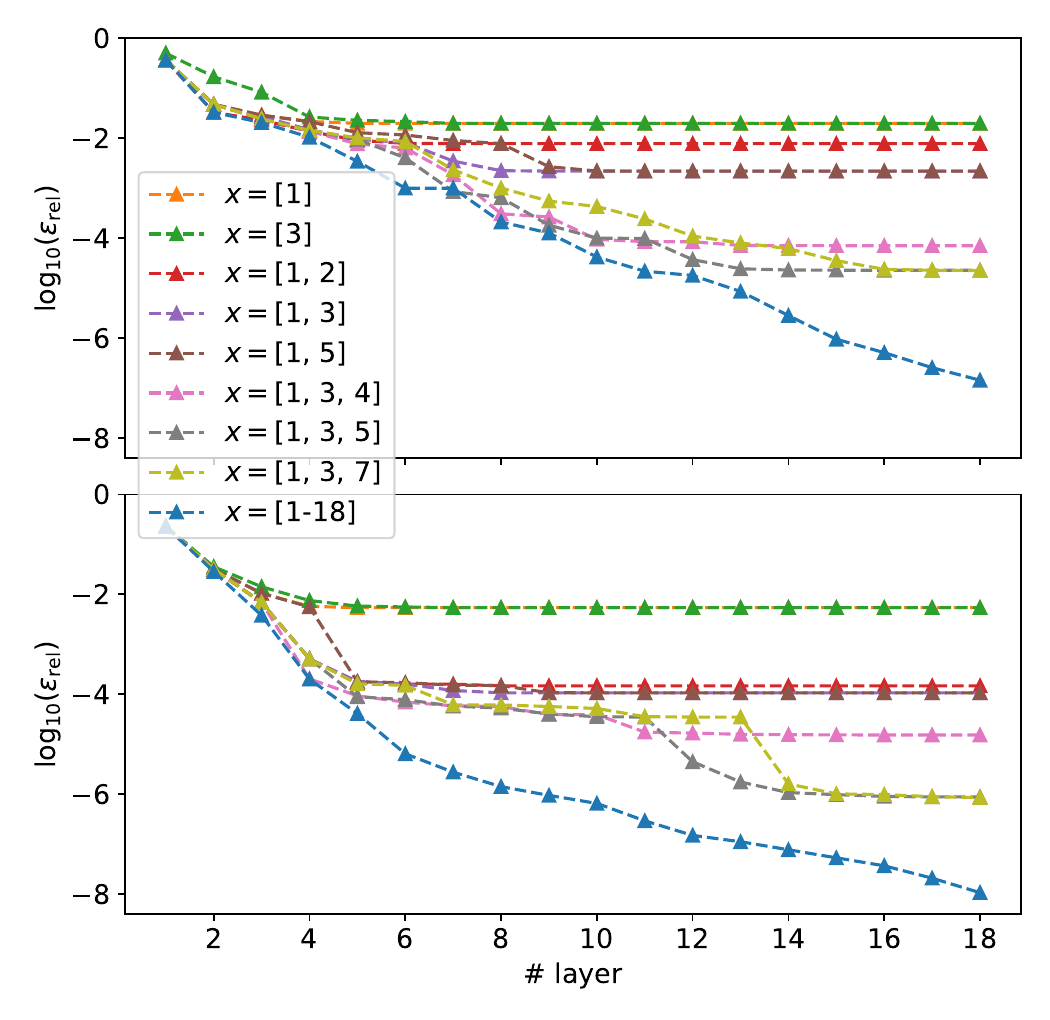}
          \vspace{-0.5em}
          \caption[]{ }
          \label{f:buildingplateausxxz}
      \end{subfigure}
      \begin{subfigure}[t]{0.48\textwidth}
          \includegraphics[width=8.5cm]{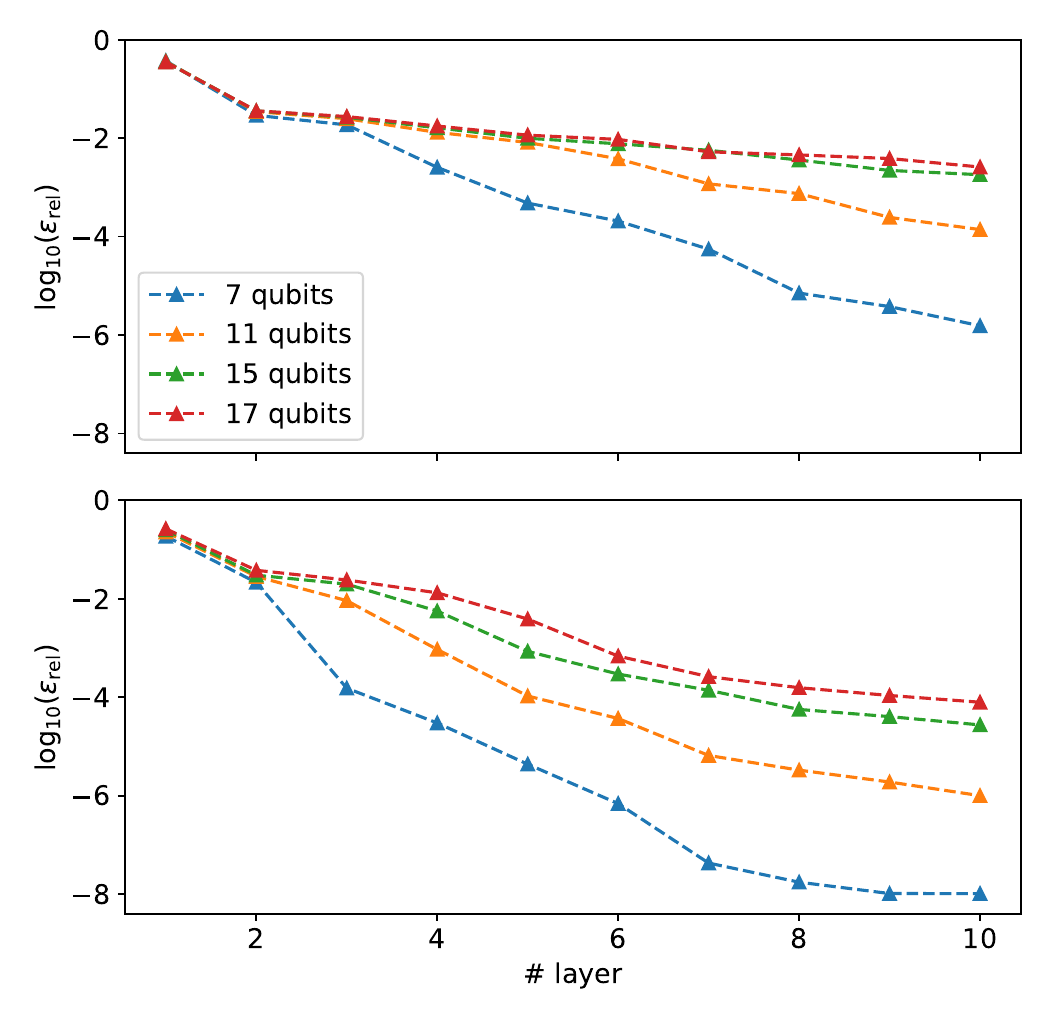}
          \vspace{-0.5em}
          \caption[]{ }
          \label{f:regimehalfedxxz}
      \end{subfigure}
      \caption{\justifying 
      (a) Accuracy obtained by the VQE for the XXZ model at the critical point without impurity, $h_0^z = 0$, (top) 
          and with impurity, $h_0^z=-10$, (bottom) for 9 qubits
          compared to the exact result in a logarithmic plot. Results for different $x$ are shown.
      (b) Accuracy depending on the number of layers for the XXZ model with $x = [\mathrm{all}]$ for 
          varying system size without impurity, $h_0^z = 0$, (top) and with impurity, $h_0^z=-10$, (bottom).
      }
    \end{figure}

    An overview of the results for $9$ qubits in the XXZ model is shown in Fig.\ \ref{f:buildingplateausxxz}.
    Overall, we find a similar behavior for the XXZ model, although less clear as for the TFIM. A key difference is in the placement of the entangling gates between the subsystems: better results are achieved when the entanglement gates are applied in every odd layer, e.g., $x = \left[ 1, 3, 5 \right]$.

    To investigate the halving of layers needed to reach exponential improvement, we compute the accuracy depending on the number of layers for $x = [\mathrm{all}]$ for various system sizes, visualized in Fig.\@ \ref{f:regimehalfedxxz}.

    Without impurity, we see again a finite-depth regime followed by a finite-size regime depending linearly on 
    the system size. Note that the distinction was very clear for the TFIM while for the XXZ model it is not as clear.
    With impurity, we find also for the XXZ model that the finite-depth regime is halved similarly our result for the TFIM.

\begin{figure}[htb]
      \centering
      \includegraphics[width=8.5cm]{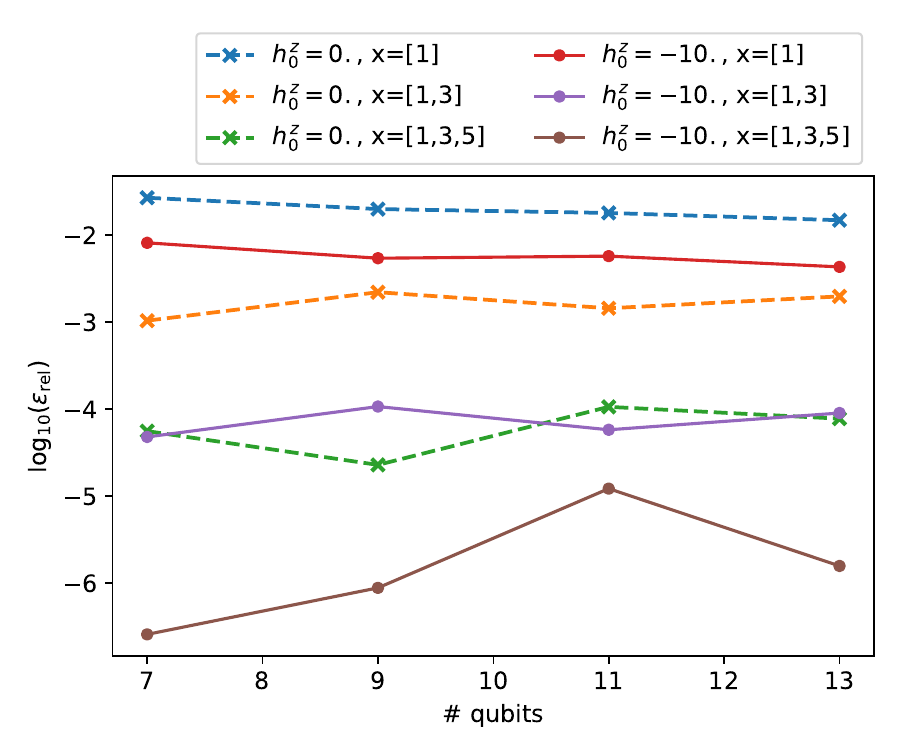}
      \vspace{-1.5em}
      \caption[]{\justifying Logarithmic accuracy obtained with the VQE compared to the exact result for different system sizes and
      number of layers in $x$ for the XXZ model at $\Delta = 0.5$. The number of $x$ also indicates the number of cuts 
      necessary to divide the circuit. Thus, one can read off the number of cuts necessary to achieve a 
      given accuracy, e.g., to achieve an accuracy in the order of $10^{-4}$ for the model with impurity two cuts are needed.}
      \label{f:readoffaccuracytxxz}
    \end{figure}

    The data of the plateau values for $x=[1], x=[1,3]$ and $x=[1,3,5]$ for increasing system sizes are plotted in Fig.\@ \ref{f:readoffaccuracytxxz}.
    The results match with those of the TFIM: Upon increasing the number of layers in $x$, the accuracy of the result increases for all chain lengths. Without impurity, the plateau values are higher than with impurity, although not as much as in the case of the TFIM. 
    Note that in general, for the XXZ model it is more difficult to achieve high precision in the minimization process and more iterations are necessary. This significantly increases the simulation time. Therefore, only system sizes up to 13 qubits are simulated.
    
\begin{figure}[htb]
      \centering
      \begin{subfigure}[c]{0.48\textwidth}
          \includegraphics[width=8.5cm]{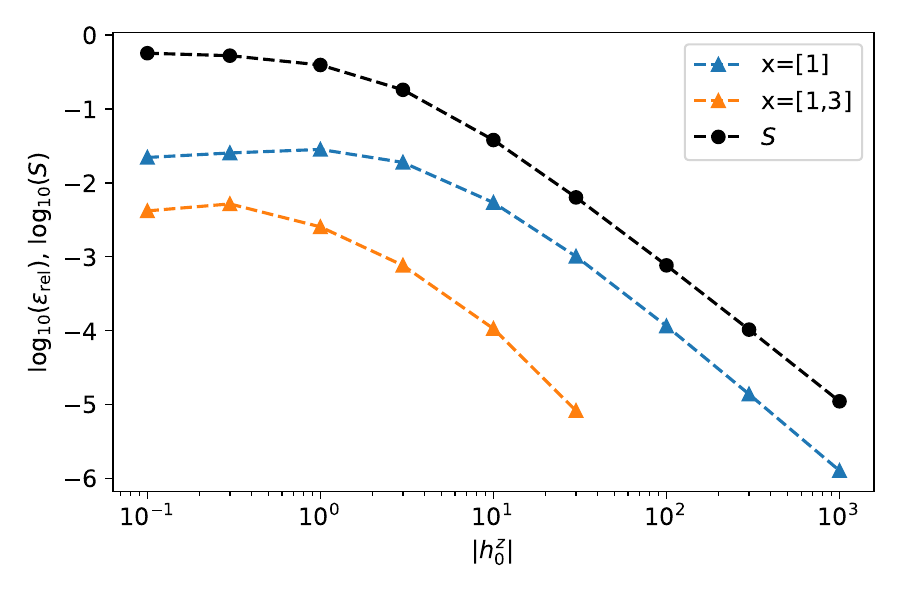}
          \vspace{-1.5em}
          \caption[]{}
          \label{f:entangentrxxz}
      \end{subfigure}
      \begin{subfigure}[c]{0.48\textwidth}
          \includegraphics[width=8.5cm]{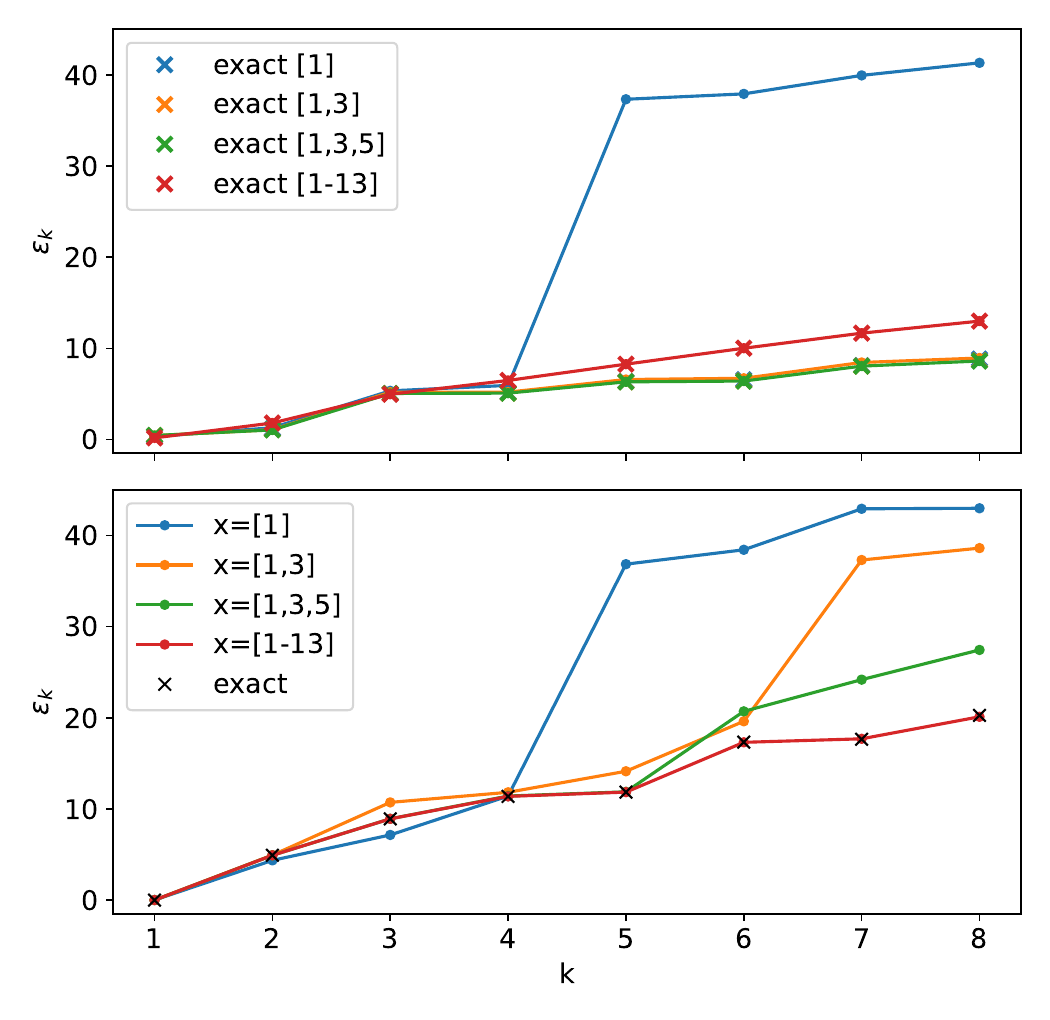}
          \vspace{-1.5em}
          \caption[]{
        }
          \label{f:esxvarxxz}
      \end{subfigure}
      \caption{\justifying 
      (a) Logarithmic accuracy obtained with the VQE at the plateau for $x=[1]$ compared to entanglement entropy for varying impurity strength for the XXZ model with 9 qubits. 
      (b) Eigenvalues $\varepsilon_k$ of the entanglement Hamiltonian for the XXZ model at $\Delta = 0.5$ with 7 qubits
          without $h_0^z = 0$, (top) and with impurity, $h_0^z=-10$, for 13 layers for different $x$ compared to the exact 
          solution. 
          In the case with impurity, different exact solutions, which are given by the largest overlap of the vector found by VQE and the linear combination of the exact calculation are present due to a degeneracy.
          This degeneracy is broken for finite $h_0^z$.
      }
      \vspace{-1em}
    \end{figure}

    To better understand why the difference of the 
    plateau values for the XXZ model is not as pronounced in the presence of an impurity as it was for the TFIM, we 
    plot the exact entanglement entropy, obtained by numerical diagonalization, as function of $h_0^z$ and the relative energy error $\varepsilon_{\mathrm{rel}}$ for $x=[1]$ and $x=[1,3]$ in Fig.\@ \ref{f:entangentrxxz}.
    With increasing magnetic field strength, the entanglement entropy
    decreases and in parallel the accuracy improves, following the same trend as for the TFIM. 
    However, much larger impurity values are needed to significantly reduce the 
    entanglement entropy. 

    In Fig.\@ \ref{f:esxvarxxz}, the entanglement spectrum is shown for the XXZ model with 7 qubits
    with and without impurity for 13 layers for different $x$.
    Without impurity, there is a degeneracy of the ground state energy, which implies that we have a linear 
    combination of potential eigenvectors $| \psi'(\vec{\theta}) \rangle$ that can be approximated by the VQE.
    Therefore, in this case we use the linear combination of the exact calculation with the largest 
    overlap with the vector found by the VQE. With this linear combination of the exact solution we calculate the exact entanglement spectrum meaning that we can have
    multiple exact spectra depending on what the largest overlap is. These are then compared to the results 
    obtained by VQE calculation.
    Using this method we find similar results as for the TFIM: The more layers are included in $x$, the more accurately the eigenvalues are found starting from the smallest ones.
    Again, with impurity the difference between the first two eigenvalues increases indicating separation of the qubit chain.
    However, the entanglement eigenvalues of the XXZ model with impurity remain small compared to those of the TFIM.
    This means that the XXZ chain is more entangled than the TFIM chain for the same value of $h_0^z$, which aligns with the results of the entanglement entropy.
     
    \end{appendix}	

\end{document}